\documentclass[a4paper,fleqn]{cas-sc}

\pdfoutput=1

\usepackage[authoryear]{natbib}
\usepackage{graphicx} 
\graphicspath{ {./Figures/} }
\usepackage{natbib}       
\usepackage{moreverb}
\usepackage[utf8]{inputenc}
\usepackage{psfrag}
\usepackage{amsmath}
\usepackage{amsthm}
\theoremstyle{plain}
\newtheorem{assumption}{Assumption}
\theoremstyle{plain}
\newtheorem{theorem}{Theorem}
\theoremstyle{definition}
\newtheorem{remark}{Remark}
\theoremstyle{definition}
\newtheorem{definition}{Definition}
\theoremstyle{definition}

\newtheorem{lemma}{Lemma}
\usepackage{amssymb}
\usepackage{amsfonts}
\usepackage{float}
\usepackage{multirow}
\usepackage{setspace}
\usepackage{flushend}
\usepackage{array}
\usepackage{tabu}
\usepackage{bm}
\usepackage{color}
\usepackage{url}
\usepackage{dsfont}
\usepackage[figuresright]{rotating}
\usepackage{subfigure}
\usepackage{svgcolor}
\usepackage{makecell}
\usepackage{booktabs}
\usepackage{scalerel}
\usepackage{eurosym}
\usepackage[dvipsnames]{xcolor}
\usepackage{footnote}
\usepackage{siunitx}
\usepackage{cancel}
\usepackage[english]{babel}
\usepackage{caption}
\usepackage[dvips]{epsfig}
\usepackage{cancel}

\newcommand{\nn}{\nonumber}
\newcommand{\setU}{\mathcal U}

\newcommand{\R}{\mathbb R}
\newcommand{\x}{\mathbf{x}}
\newcommand{\y}{\mathbf{y}}
\newcommand{\G}{\mathbf{G}}
\renewcommand{\u}{\mathbf{u}}
\newcommand{\utar}{u_\mathrm{tar}}

\newcommand\BibTeX{{\rmfamily B\kern-.05em \textsc{i\kern-.025em b}\kern-.08em
		T\kern-.1667em\lower.7ex\hbox{E}\kern-.125emX}}
\def\tsc#1{\csdef{#1}{\textsc{\lowercase{#1}}\xspace}}
\tsc{ASV}
\tsc{USV}
\tsc{ASC}
\tsc{ISS}

\begin{document}

\let\WriteBookmarks\relax
\def\floatpagepagefraction{1}
\def\textpagefraction{.001}
\shorttitle{Nonlinear model predictive control-based guidance law for path following of unmanned surface vehicles}
\shortauthors{G. Bejarano et~al.}

\title [mode = title]{Nonlinear model predictive control-based guidance law for path following of unmanned surface vehicles\footnote{© 2022. This manuscript version is made available under the CC-BY-NC-ND 4.0 license \url{https://creativecommons.org/licenses/by-nc-nd/4.0/}. The link to the formal publication is \url{https://doi.org/10.1016/j.oceaneng.2022.111764}}}                      
\tnotemark[1]

\author[1]{Guillermo Bejarano}[type=editor,
                        auid=000,bioid=1,
                        role=, orcid=0000-0002-2951-8829]
\ead{gbejarano@uloyola.es}                        
\author[1]{José María Manzano}[auid=000,bioid=2,role=,orcid=0000-0003-2932-5581]
\cormark[1]
\ead{jmanzano@uloyola.es}

\author[1]{José Ramón Salvador}[type=editor,
                        auid=000,bioid=3,
                        role=, orcid=0000-0002-1568-1318]

\ead{jrsalvador@uloyola.es}

\author[2]{Daniel Limon}[%
   role=,orcid=0000-0001-9334-7289,
   ]
\ead{dlm@us.es}

\address[1]{Department of Engineering, Universidad Loyola Andaluc{\'\i}a, Avenida de las Universidades s/n. 41704 Dos Hermanas, Sevilla, Espa\~na}

\address[2]{Systems Engineering and Automation Department. Universidad de Sevilla, Camino de los Descubrimientos, s/n. 41092 Sevilla, Espa\~na}

\cortext[cor1]{Corresponding author}

\begin{abstract}
This work proposes a nonlinear model predictive control-based guidance strategy for unmanned surface vehicles, focused on path following. The application of this strategy, in addition to overcome  drawbacks of previous line-of-sight-based guidance laws, intends to enable the application of predictive strategies also to the low-level control, responsible for tracking the references provided by the guidance strategy. The stability and robustness of the proposed strategy are theoretically discussed. Furthermore, given the non-negligible computational cost of such nonlinear predictive guidance strategy, a practical nonlinear model predictive control strategy is also applied in order to reduce the computational cost to a great extent. The effectiveness and advantages of both proposed strategies over other nonlinear guidance laws are illustrated through a complete set of simulations.
\end{abstract}


\maketitle

\section{Introduction}\label{sec_intro}

Unmanned surface vehicles (USVs), also called autonomous surface vehicles (ASVs), have drawn increasing attention in the last years for several applications in ocean space \citep{shi2017advanced}. USVs are able to perform autonomously a wide variety of operations in challenging marine and coastal environments, avoiding direct human intervention and thus all associated risks. Moreover, they can even address tasks at locations otherwise unreachable for manned vessels. Marine science, ocean oil and gas exploration, offshore renewables, and border surveillance, among other scientific and military applications, have boosted the interest for this type of vessels \citep{liu2016unmanned,tanakitkorn2019review}. Among all types of USVs, small-size and low-cost vessels have some particular advantages when compared to larger crafts, such as lighter weight, modularity, and increased maneuverability and versatility \citep{wynn2014autonomous,zhang2015future}.

Robust and effective USV motion control is a key problem to be addressed when facing fully autonomous tasks, especially in close-range operating areas or shallow waters \citep{li2019toward}. Indeed, this problem has drawn the interest of the control community in the last decades, since model nonlinearity, unmodelled or roughly estimated hydrodynamics, parametric model uncertainty, and non-measurable disturbances convert it into a challenging problem \citep{breivik2010topics}.

In the literature there are mainly three different approaches for USV motion control: dynamic positioning, trajectory tracking, and path following (PF) control  \citep{aguiar2005,LAPIERRE20071734}. The first one seeks to keep the position and heading of the USV within predefined limits, while the second one tries to track a spatial trajectory with hard time constraints. The PF approach is similar to trajectory tracking, but the time constraints of the path to be followed are much looser. This makes PF more suitable for practical situations with time-varying currents, where trajectory tracking could require highly demanding actuator performance, or could be even unfeasible.

\begin{figure}[!ht]
	\begin{center}
		\includegraphics[width=0.8\textwidth,trim=20 210 360 120,clip]{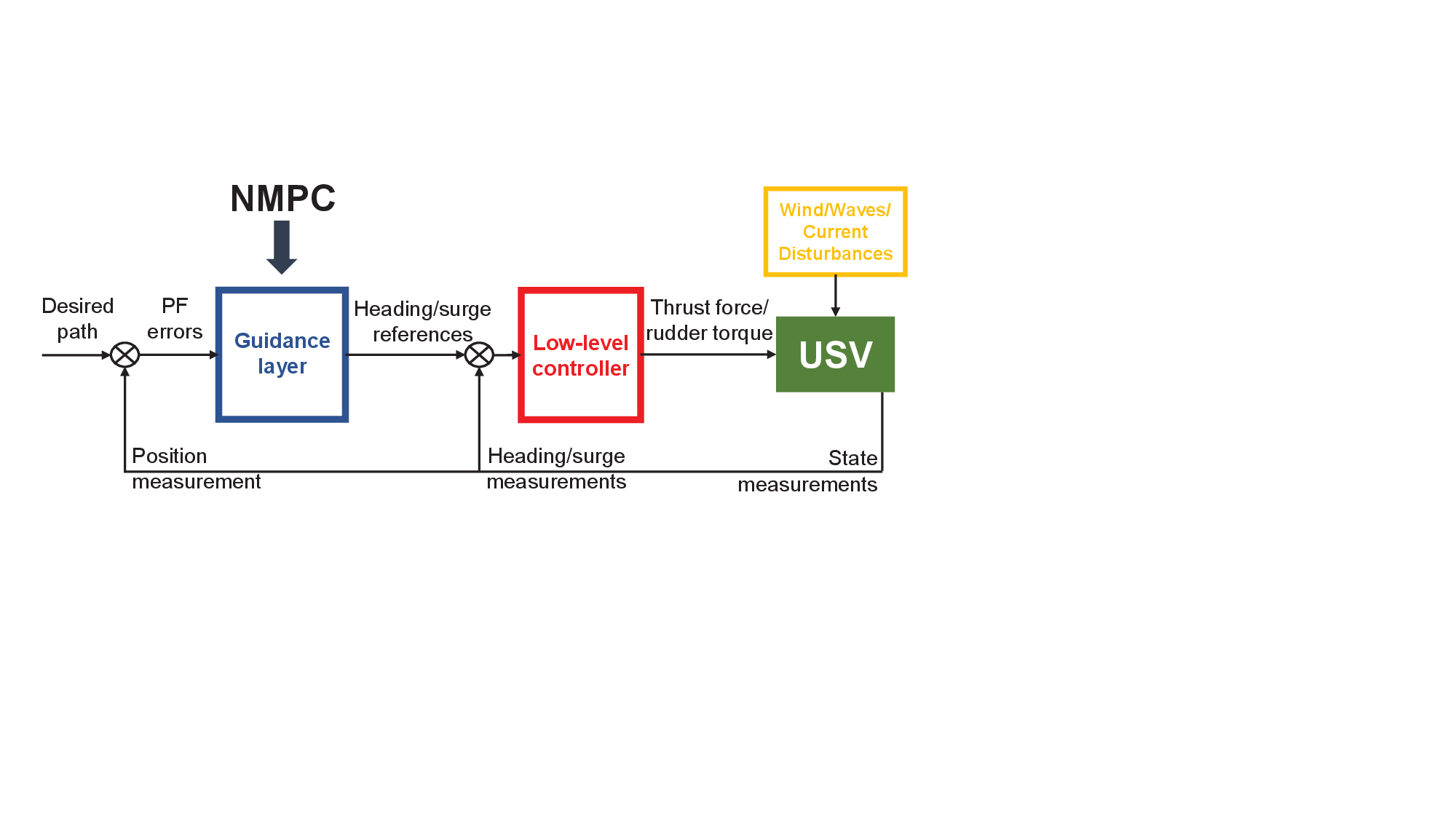}
		\caption{Traditional PF control strategy} 
		\label{figEstimationControlStrategy}
	\end{center}
\end{figure}

The PF strategies are traditionally divided into two layers in a cascade structure \citep{fossen2003line,bejarano2020velocity}, as indicated in  Fig.~\ref{figEstimationControlStrategy}. On the one hand, the high-level layer, also known as the guidance layer, is responsible for generating the set points for the heading angle and forward speed, as well as their respective time derivatives, in such a way that the desired path is followed and the time constraints concerning the desired forward speed are met. On the other hand, the low-level layer includes a controller that manipulates the actuators in such a way that the set points provided by the guidance law are tracked, despite non-measurable wind, waves, and current disturbances, in addition to high model uncertainty. In typical underactuated USV configurations (usually preferred for high-speed maneuvering in long-range and long-duration missions due to cost-effective and weight considerations), the actuator signals are only the thrust force and the rudder torque. Indeed, even fully-actuated USVs equipped with lateral actuators decrease dramatically their
efficiency in lateral directions at high-speed forward movement \citep{xiang2015smooth}. Then, the thrust force and the rudder torque must be translated into the propeller velocities, using the propeller models and the specific layout of the USV. Therefore, in the traditional PF strategy, the guidance law focuses on the kinematic problem, while the low-level controller focuses on the kinetic problem.

Many different control strategies have been proposed for the low-level control, seeking to reject disturbances and compensate for model uncertainty: parametric adaptive methods \citep{skjetne2005adaptive}, robust control \citep{lekkas2014integral}, sliding-mode control \citep{CUI201645,chen2017sliding}, and fuzzy logic systems and neural networks \citep{wang2019fuzzy}, among many others. However, regarding the guidance law, the well-known line-of-sight (LOS) law \citep{breivik2005guidance,fossen2011handbook} and its varieties have been proven effective when following a straight-line or curve path in absence of ocean currents and disturbances. This technique, based on how an experienced sailor would steer the helm, has been enhanced over the years to deal with disturbances and time-varying currents \citep{wang2020guidance}. Integral LOS \citep{borhaug2008integral,lekkas2014integral}, adaptive LOS \citep{fossen2015line,zeng2017adaptive}, compound LOS \citep{miao2017compound}, and extended state observer-based LOS \citep{liu2016eso,yu2019elos} are all strategies based on the original LOS that try to estimate and then compensate the sideslip angle caused by time-varying disturbances. Recently, variants of the LOS law have been also applied to automatic steering of sailboats \citep{deng2020line}. Different references for the heading angle are proposed for the so-called path following, tacking, and gybing modes, based on the double reduced-order extended state observer estimation of the crab angle, which plays the role of the sideslip angle defined in common ships. 

However, in all the mentioned works the kinematic problem is somehow decoupled into two parts: (1) to drive the PF errors to zero, by computing the set point for the heading angle, and (2) to make the reference for the USV forward speed or surge match a predefined value. Using this structure, the kinematic problem is indeed solved by using the rudder torque as the only control action, since the thrust force is completely devoted to ensuring the surge reference tracking. \cite{wang2019fuzzy} have recently proposed a surge-guided LOS (SGLOS) where the set point for the surge is modified according to the PF errors, giving rise to enhanced USV manoeuvrability. By avoiding decoupling, both actuation signals collaborate to get the USV to follow the desired path. When the vessel is far away from the desired trajectory, the surge reference is higher than the desired value and it eventually converges to the predefined value as the PF errors converge to zero and the USV approaches the desired path.  

The original LOS strategy and all its varieties are shown to be effective and easy-to-implement guidance laws. However, in the typical case when the desired path and surge are a priori known, no benefit is taken from this knowledge. Furthermore, since the set points for the low-level control are computed by the guidance law at every sampling time and no future references are computed, no model predictive control (MPC) strategies are usually applied at the low-level control. In spite of the disturbances being non-measurable and unpredictable, if the future references were available, the performance of the low-level controller could be still enhanced using predictive strategies. 

In this work, due to the previously remarked drawbacks of LOS-based guidance laws, specifically the SGLOS \citep{wang2019fuzzy}, a guidance law based on nonlinear MPC (NMPC) is proposed. The main contributions of this paper yield certain advantages of the proposed predictive law over the SGLOS~\cite{wang2019fuzzy}, which are summarised as follows:
\begin{itemize}
	\item The {a priori} knowledge of the desired path and surge is used to drive the PF errors to zero in a more efficient way, thanks to the predictive feature.
	\item Even when the USV is far away from the desired path, a desired surge can be imposed.
	\item Constraints on the rate of change of the surge and heading references can also be imposed, in order to account for the low-level controller and actuator dynamics.
	\item An existing robust and stable control law is tailored to the path following problem.
	\item A fast practical implementation of the nonlinear model predictive control law is also proposed.
\end{itemize}
In addition, although this work does not exploit them, the proposed predictive law enables other possible advantages:
\begin{itemize}
	\item {Constraints on the PF errors may be considered in the optimization problem, enabling obstacle avoidance if their position were described in terms of the desired trajectory.} 
	\item {Computed future references for the heading and surge may enhance tracking performance, if a predictive low-level controller is applied.}
\end{itemize}

The model predictive PF problem has also been addressed by the control community. However, these controllers typically merge the two different control layers that are worth considering in USV applications: the guidance problem and the reference tracking (see Fig.~\ref{figEstimationControlStrategy}). For instance, \cite{faulwasser2009model} propose a stabilizing NMPC for the PF problem, providing nominal stability guarantees of the closed-loop system. \cite{alessandretti2013trajectory} study both the trajectory tracking and the PF problems in terms of NMPC, again considering a single dynamical system. Hence, desired surge requirements are not addressed in these PF problems. For this reason, a multiobjective MPC for PF was recently proposed by~\cite{shen2018path}, aiming to guarantee path convergence while being able to drive the surge to a given desired value. Because of the guidance system considered in this paper, the speed assignment will be treated here as an input reference, avoiding the need for a multiobjective MPC.

Besides, we take advantage of a formulation of the MPC problem that weights the terminal cost~\citep{limon2006stability}, avoiding the need for a terminal constraint, tailored in this paper to the PF scheme. This lifts the cumbersome calculation of a terminal set, in contrast to aforementioned approaches, in which e.g., the USV is required to reach the reference at the end of the prediction horizon~\citep{faulwasser2009model,alessandretti2013trajectory,shen2018path}.

Moreover, a robust design is proposed, taking into account uncertainty from unmodelled lateral dynamics and due to currents, waves, and wind. In this respect, input-to-state stability~\citep{limon2009input} of the closed-loop signal is achieved, which is, to the best of the authors' knowledge, another novel contribution in the context of PF problems.

One of the major disadvantages of the nonlinear strategies is often related to computational cost, when compared to the easy-to-implement and effective LOS-based strategies. The main reason behind high computational costs lies in the nonlinear features of the path-following-error model and the associated optimization problem, which might hinder the application of NMPC strategies in low-cost USVs. 

In this paper, the proposed NMPC-based guidance law is also simplified to reduce the computational burden, using the main idea of the Practical NMPC (PNMPC) technique \citep{plucenio2007practical,Plucenio}. This technique deals with nonlinear systems using the MPC techniques developed for linear systems and thus extremely fast to implement. A linear representation of the predicted output with regard to the future increments of the control actions is intended, by means of a linearization along the trajectory. A first-order linearization based on the corresponding Jacobian matrix is recomputed every sampling time, giving rise to a linearized model around the current point and a linear implementation of the optimization procedure, which reduces the computational cost to a great extent.

The remainder of the work is organised as follows. Section~\ref{secModellingProblemFormulation} presents the USV modelling and formulates the PF problem. The proposed NMPC-based guidance law is presented in Section~\ref{secNMPC}, whereas Section~\ref{secPNMPC} describes the application of the PNMPC technique to reduce the computational cost of the previously proposed guidance law. Section~\ref{secSimulation} provides some illustrating simulation results, comparing the performance of the proposed guidance law to existing strategies and showing the computational cost reduction provided by the linearized PNMPC-based guidance law. Finally, Section~\ref{secConclusions} summarises the main conclusions and expresses some future work.


\section{Problem formulation} \label{secModellingProblemFormulation}

\begin{figure}[!ht]
	\begin{center}
		\includegraphics[width=9.5cm,trim = 140 85 290 65, clip]{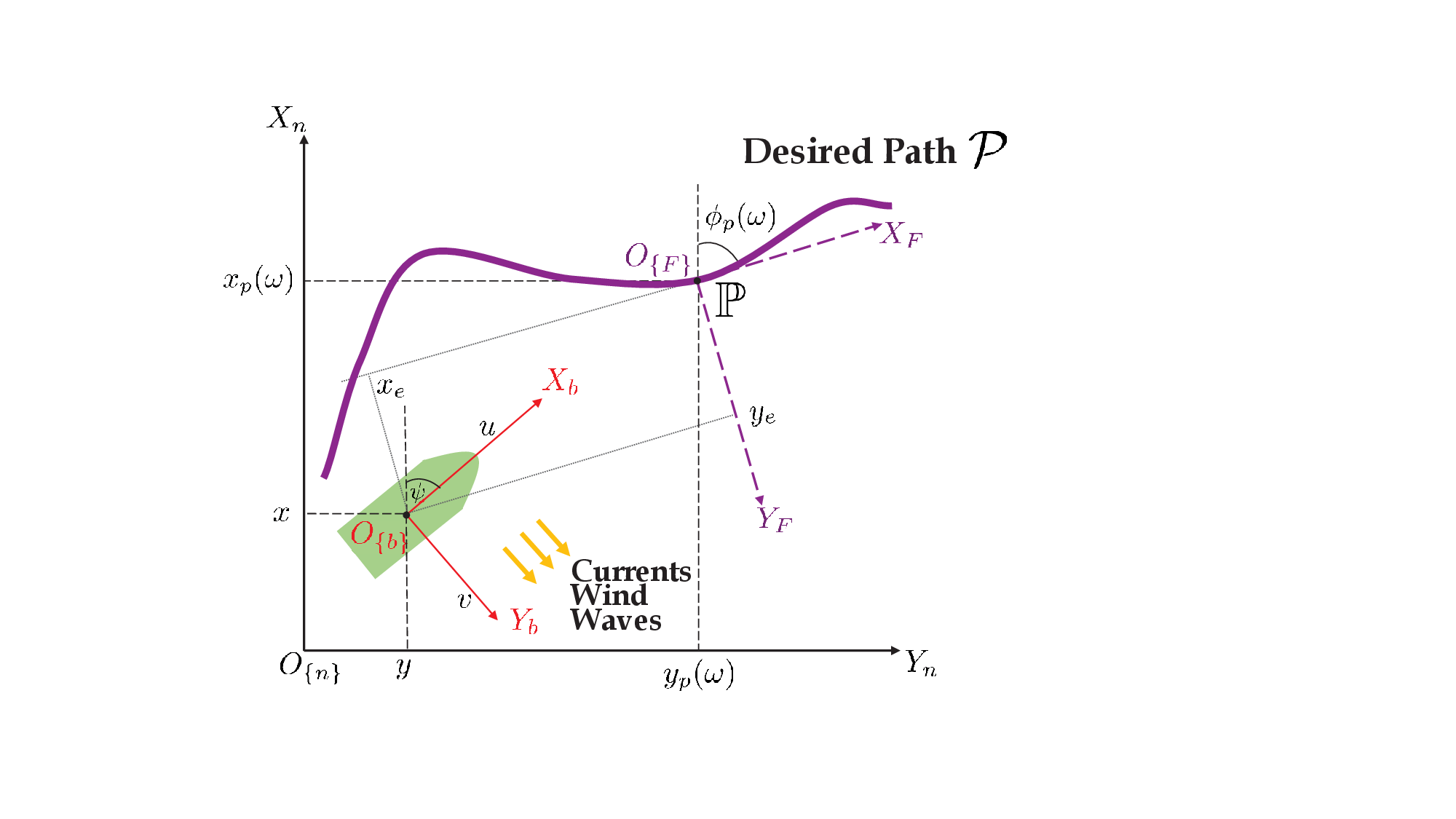}
		\caption{Path following geometry}
		\label{figPFC}
	\end{center}
\end{figure}

The three-degree-of-freedom kinematic and kinetic model of an USV moving in a horizontal plane is given by Eq.~\eqref{eqFossenModel}, according to~\cite{fossen2011handbook}:
\begin{equation}
	\begin{aligned}
		\left\{ 
		\begin{matrix}
			\begin{aligned}
				\dot{\bm{\eta}} &= \bm{R}(\psi) \bm{\nu} \\
				\bm{M} \dot{\bm{\nu}} &= - \bm{C}(\bm{\nu})\bm{\nu} - \bm{D}(\bm{\nu})\bm{\nu} - \bm{g}(\bm{\eta}) + \bm{\tau_{w}} + \bm{\tau} , \\
			\end{aligned}
		\end{matrix}
		\right.
	\end{aligned}
	\label{eqFossenModel}
\end{equation}
where $\bm{\eta} = [x \quad y \quad \psi]^T$ includes the planar vessel position and heading expressed in an earth-fixed inertial frame \{n\}, $\bm{\nu} = [u \quad v \quad r]^T$ gathers the surge, sway, and yaw velocities expressed in the body-fixed frame \{b\}, $\bm{\tau} = [F_u \quad 0 \quad \tau_r]^T$ represents the control action vector, and $\bm{\tau}_{w} = [ F_{w,u} \quad F_{w,v} \quad \tau_{w,r} ]^T$ contains the corresponding environmental forces and torque due to wind, waves, and currents (see Fig.~\ref{figPFC}). Notice that an underactuated configuration is considered, since only $F_u$ and $\tau_r$ are available as control actions, without loss of generality.

The rotation matrix $\bm{R}(\psi)$ between the body-fixed frame \{b\} and the earth-fixed inertial frame \{n\}, as defined by \cite{fossen2011handbook}, is expressed in Eq.~\eqref{eqMatrixR}, where $\bm{R}_2(\psi) \in \mathbb{R}^{2\times2}$ is the upper left two-dimensional submatrix of the original rotation matrix $\bm{R}(\psi)$. $\bm{M} = \bm{M}^{T}$ is the inertia matrix, $\bm{C}(\bm{\nu})$ refers to the Coriolis/centrifugal matrix, $\bm{D}(\bm{\nu})$ to the damping matrix, and $\bm{g}(\bm{\eta})$ represents the gravitational and buoyancy forces and torques, according to the nomenclature defined by \cite{skjetne2004nonlinear}.
\begin{equation} 
	\bm{R}(\psi) = \left[
		\begin{matrix}
			\text{cos}(\psi) & -\text{sin}(\psi) & 0 \\
			\text{sin}(\psi) &  \text{cos}(\psi) & 0 \\
			0		  &	 0		   & 1 \\
		\end{matrix}
	\right] = 
	\left[
	\begin{matrix}
		\bm{R}_2(\psi) & \bm{0} \\
		\bm{0}		   & 1 \\
	\end{matrix}
	\right]
	\label{eqMatrixR}
\end{equation}

The desired planar path $\mathcal P$, as represented in Fig.~\ref{figPFC}, is assumed to be available {a priori} and parameterised by a time-dependent path variable $\omega (t)$, such that:
\begin{equation}
	\mathcal P = \{\mathbb P (\omega)\in\mathbb R^2 : \omega\in[0,\infty)\mapsto \mathbb P(\omega) \equiv [x_p(\omega) \quad y_p(\omega)]^T\}.
\end{equation}

The position of the moving virtual target $\mathbb P$ on the path is assumed to be defined by the time-varying value of the path variable $\omega$. From the virtual target $\mathbb P$, the Frenet-Serret frame \{F\} \citep{Serret1851,Frenet1852} can be defined, orthogonally to the path, as represented in Fig.~\ref{figPFC}. The velocity of the virtual target, $\utar$, is related to the time derivative of~$\omega$ through Eq.~\eqref{eqVirtualTargetVelocity}:
\begin{equation}
	\begin{aligned}
		\utar &= \sqrt{\dot{x}_p^2 + \dot{y}_p^2} =  \dot{\omega} \, \sqrt{(\partial x_p / \partial \omega)^2 + (\partial y_p / \partial \omega)^2}.\\
	\end{aligned} 
	\label{eqVirtualTargetVelocity}
\end{equation}
Furthermore, in order to prove robust stability of the closed-loop system, a signal that tends to zero as~$\omega\to\infty$ needs to be introduced. For example, the variable $z$ as a function of $\omega$ is defined as indicated in Eq.~\eqref{eqZDefinition}:
\begin{equation}
	z (\omega) \equiv \dfrac{1}{\omega+1}.
	\label{eqZDefinition}
\end{equation}
Applying the chain rule, Eq.~\eqref{eqVirtualTargetVelocity} can be also expressed as indicated in Eq.~\eqref{eqzpunto}:
\begin{equation}
	\dot z = \dfrac{-1}{(\omega+1)^2}\dot\omega = \dfrac{-z^2 \utar}{F(\omega)},
	\label{eqzpunto}
\end{equation}
\noindent where~$F(\omega)$ is a known function of the virtual target~$\mathbb P$ given the desired path~$\mathcal P$, as shown in Eq.~\eqref{eqFomega}:
\begin{equation}
	F(\omega) = \sqrt{(\partial x_p / \partial \omega)^2 + (\partial y_p / \partial \omega)^2}.
	\label{eqFomega}
\end{equation}
The angle $\phi_p (\omega)$ between the Frenet-Serret \{F\} and the earth-fixed \{n\} frame is defined by Eq.~\eqref{eqGammap}:
\begin{equation}
	\phi_p (\omega) = \mathrm{atan2}(\dot{y}_p,\dot{x}_p) = \mathrm{atan2}(\partial y_p / \partial \omega,\partial x_p / \partial \omega),
	\label{eqGammap}
\end{equation}
\noindent where it should be noticed that $\mathrm{atan2}(y,x)$ refers to the~$\arctan(y/x)$ function extended to the four quadrants.
The PF errors between the planar position of the vessel, ($x$, $y$), and the position of the virtual target $\mathbb P$, ($x_p(\omega)$,~$y_p(\omega)$), expressed in \{F\}, are defined by Eq.~\eqref{eqPFCErrors}:
\begin{equation}
	\begin{aligned}
		\left[
		\begin{matrix} 
			x_e \\
			y_e \\
		\end{matrix}
		\right]
		&\equiv \bm{R}_2^{T}(\phi_p) \cdot
		\left[
		\begin{matrix} 
			x - x_p(\omega) \\
			y - y_p(\omega) \\
		\end{matrix}
		\right]
		= \left[
		\begin{matrix}
			\text{cos}(\phi_p) & -\text{sin}(\phi_p) \\
			\text{sin}(\phi_p) &  \text{cos}(\phi_p) \\
		\end{matrix}
		\right]^{T} \cdot
		\left[
		\begin{matrix} 
			x - x_p (\omega) \\
			y - y_p (\omega) \\
		\end{matrix}
		\right].
	\end{aligned}
	\label{eqPFCErrors}
\end{equation}
$x_e$ and $y_e$ are widely known as the along- and cross-track errors, respectively. Hence, deriving Eq.~\eqref{eqPFCErrors} with respect to time and introducing Eqs.~\eqref{eqVirtualTargetVelocity}~and~\eqref{eqFomega}, the PF error dynamics indicated in Eq.~\eqref{eqPFCErrorDynamics} are obtained: 
\begin{equation}
	\begin{aligned}
		\dot{x}_e &= u \, \cos(\psi - \phi_p) - v \, \sin(\psi - \phi_p) + \dot{\phi}_p \, y_e - \utar  \\
		&= u \, \cos(\psi - \phi_p) - v \, \sin(\psi - \phi_p) + \utar \, \left( \dfrac{1}{F(\omega)} \, \dfrac{\partial \phi_p}{\partial \omega} \, y_e  - 1 \right), \\
		\dot{y}_e &=  u \, \sin(\psi - \phi_p) + v \, \cos(\psi - \phi_p) - \dot{\phi}_p \, x_e  \\
		&= u \, \sin(\psi - \phi_p) + v \, \cos(\psi - \phi_p) - \utar \, \dfrac{1}{F(\omega)} \, \dfrac{\partial \phi_p}{\partial \omega} \, x_e.  \\
	\end{aligned} 
	\label{eqPFCErrorDynamics}
\end{equation}

Eq.~\eqref{eqPFCErrorDynamics}, together with the definition of~$\dot z$ in Eq.~\eqref{eqzpunto}, represents the dynamics of the system to be controlled by the NMPC-based guidance law described in Section \ref{secNMPC}, given that~$x_p$, $y_p$, $\phi_p$, and~$F$ are assumed to be known functions of~$z$, or conversely~$\omega$.

\section{Nonlinear model predictive control} \label{secNMPC}
From this section onwards, the following notation will be used.

\subsection*{Notation}
Given two column vectors,~$v$ and $w$,~$(v,w)$ stands for~$[v^T,w^T]^T$. The set of integer numbers from~$a$ to~$b$ is denoted~$\mathbb I_a^b$. A continuous function~$\alpha:\R_{\geq 0}\to\R_{\geq 0}$ is a~$\mathcal K$-function if it is strictly increasing and~$\alpha(0)=0$. Besides, if~$\alpha$ is unbounded, it is called a~${K_\infty}$-function. A continuous function~$\beta:\R^2_{\geq 0}\to\R_{\geq 0}$ is a~$\mathcal {KL}$-function if~$\beta(s,t)$ is~$\mathcal K_\infty$ in~$s$ for all~$t$ and~$\lim_{t\to\infty}\beta(s,t)=0$ for all~$s\geq 0$. A ball of radius~$r\in\R^n$ is the set~$\mathcal B(r)=\{x:|x|\leq r\}$, where~$|x|$ is the component-wise absolute value of the vector. The weighted norm of a vector~$v\in\R^{n}$ with coefficient~$\bm{M}\in\R^{n\times n}$ is denoted~$\|v\|_M = v^T\bm{M}v$. The predicted values for a signal~$s(t)$ are denoted~$\hat s(t)$.

\subsection{Problem statement}

The dynamical system controlled by the proposed NMPC is obtained from discretizing the continuous-time model presented in Eqs.~\eqref{eqzpunto}~and~\eqref{eqPFCErrorDynamics}, using the forward Euler approach. It is a partially virtual system, since it is extended to consider the virtual target point in the path, adding an additional degree of freedom, as described by~\cite{wang2019fuzzy}. Hence, the discrete-time model of the system to be controlled can be expressed as indicated in Eq.~\eqref{eq_model}:
\begin{equation} 
	\x(k+1)=f(\x(k),\u(k),v(k)).
	\label{eq_model}
\end{equation}
\noindent Here, the state~$\x$ is comprised by the PF errors and the alternative path variable~$z$. The input~$\u$ is given by the references for the underactuated USV (surge and heading angle), and extended to include the virtual target velocity~$\utar$. The only disturbance is assumed to be the sway velocity~$v$, which accounts for the effect of currents, waves, wind, and unmodelled dynamics, called \emph{drift forces} in the related literature~\citep{fossen2015line}.
A vessel exposed to \emph{drift forces} (currents, wind, and waves) presents variations in the velocities $u$, $v$, and~$r$ due to motion kinetics (see Eq.~\eqref{eqFossenModel}). The response can be observed as a nonzero sideslip angle $\beta \equiv \text{atan2}(v,u)$, which implies a non-zero sway velocity $v$. Thus,
\begin{eqnarray}
	\x(k)&=&(
	x_e(k),\,y_e(k),\,z(k)),\\
	\u(k)&=&(u(k),\,\psi(k),\,\utar(k)).
\end{eqnarray}
Note that the model function $f$ is hence given by Eqs.~\eqref{eqzpunto} and~\eqref{eqPFCErrorDynamics}, once discretized. In order to derive a NMPC strategy for this virtual system, the desired path must satisfy the following assumption.

\begin{assumption}\label{ass_path}
	The desired path~$\mathcal P$ is smooth, and such that~$f$ is Lipschitz continuous.
\end{assumption}

In practice, the path, which is designed a priori, can in general be chosen such that Assumption~\ref{ass_path} is satisfied. For example,~\cite{fossen2015line} require the path to be $\mathcal{C}^1$ differentiable.

Moreover, in contrast to existing LOS-based guidance laws (see e.g.~\cite{wang2019fuzzy}), the MPC framework presented in this paper allows to consider constraints. Therefore, we will impose both limits on the surge velocity and on the variation of the surge and the heading references, such that they are followable by the low-level controller. Moreover, imposing~$\utar>0$ ensures forward motion along the path, as it is customary in PF problems. Thus, the closed-loop system must satisfy:
\begin{subequations}\allowdisplaybreaks
\label{eq:constraints}
	\begin{eqnarray}\allowdisplaybreaks
		\u(k)\in\setU=\begin{Bmatrix}\u:\begin{bmatrix}
				0\\-\pi\\\epsilon
			\end{bmatrix}\leq\u\leq \begin{bmatrix}
				\bar u\\\pi\\\bar{u}_\mathrm{tar}
			\end{bmatrix}
		\end{Bmatrix},\forall k,\\
		\nonumber\\
		\Delta \u \equiv \u(k)-\u(k-1)\in\setU_g=\begin{Bmatrix}\u:|\Delta\u|\leq
			\begin{bmatrix}
				\overline{\delta u}\\\overline{\delta \psi}\\\infty
			\end{bmatrix}
		\end{Bmatrix},\forall k,
	\end{eqnarray}
\end{subequations}
\noindent with~$\epsilon>0$ to ensure forward motion, and~$\bar{(\cdot)}\in\R>0$ represents the upper bounds.

Furthermore, the MPC strategy also enables the assignment of a reference to all states (where it is trivially assumed that the PF errors are desired to converge to~$0$) and inputs. 

Next, the optimization problem of the NMPC-based guidance law is introduced:
\begin{subequations} \label{eq_opti}
	\begin{eqnarray}
		\min_{\mathbf{u}} && J_N(\x(k),\mathbf{u})\\
		&&=\sum\limits_{j=0}^{N-1}\ell(\hat \x(j|k),\u(j))+\lambda V_f(\hat\x(N|k)),\nn\\
		\mathrm{s.t.}&& \hat \x(0|k)=\x(k),\\
		&&\hat \x(j+1|k)= f(\hat \x(j|k),\u(j),v(k)),\,j\in\mathbb I_0^{N-1},\nn \label{eq_prediction_model}\\
		&&\\
		&& \u(j)\in\setU,\,j\in\mathbb I_0^{N-1},\\
		&& \u(j)-\u(j-1)\in\setU_g,\,j\in\mathbb I_0^{N-1}.\label{eq:ug}
	\end{eqnarray}
\end{subequations}
In this problem,~$J_N$ is the cost to be minimized, predicted over a horizon~$N$, composed by the stage cost~$\ell$ and the terminal cost~$V_f$. The latter is weighted by a design parameter~$\lambda\geq 1$, which allows to avoid the need for a terminal constraint~\citep{limon2006stability}. Besides, notice that, lacking a description or estimate of the disturbance, the predicted states are computed with the last available measured value,~$v(k)$.

Hence, given the measurement of the state at the current step~$\x(k)$, future predictions are obtained for the following~$N$ steps, denoted~$\hat \x(j|k)$,~$j=0,\ldots,N-1$. The optimal sequence resulting at a time~$k$ is denoted~$\bm u^*$, and its cost~$J_N^*$.

\subsection{Stability}

In order to prove the stability and convergence to the path of the closed-loop system, some standard assumptions in the design of model predictive controllers must hold true (see~\cite{rawlings2017model}).

\begin{assumption}\label{ass_stagecost}
	The stage cost $\ell$ is continuous and positive definite. Besides, given two~$\mathcal K$-function~$\alpha_x$, $\alpha_u$, then
	\begin{equation}
		\ell(\x,\u)\geq \alpha_x(\|\x\|)+\alpha_u(\|\u-\u_r\|),
	\end{equation}
	\noindent where~$\u_r$ stands for the reference of the control input.
\end{assumption}

\begin{assumption}\label{ass_terminalcost}
	There exists a terminal controller~$\kappa_f(\x)$, a terminal cost~$V_f(\x)$, and a region~$\Omega_\gamma=\{\x:V_f(\x)\leq\gamma\}$
	such that, for all~$\x\in\Omega_\gamma$:
	\begin{subequations}
		\begin{eqnarray}
			\alpha_1(\|\x\|)\leq& V_f(\x)&\leq \alpha_2(\|\x\|),\\
			V_f(\x^+)-V_f(\x)&\leq&-\ell(\x,\kappa_f(\x)-\u_r),\label{eq:deltaVf}\\
			\kappa_f(\x)&\in&\setU,\\
			\kappa_f(\x^+)-\kappa_f(\x)&\in&\setU_g,\label{eq:kug}
		\end{eqnarray}
	\end{subequations}
	\noindent where~$\alpha_1$ and~$\alpha_2$ are~$\mathcal K$-functions, and~$\x^+=f(\x,\kappa_f(\x),0)$.
\end{assumption}

Let define the so-called feasibility region~$X_N(\lambda)$, given a constant~$\varphi>0$:
\begin{equation}\label{eq_dda}
	X_N(\lambda)=\lbrace \x:J_N^*(\x(k))\leq N\varphi+\lambda\gamma\rbrace.
\end{equation}

Next, before taking into account the robust stability of the design, nominal stability (i.e., in absence of disturbances) is stated.

\begin{theorem}\label{th_nominalstab}
	Consider that Assumptions~\ref{ass_path}-\ref{ass_terminalcost} hold. Then, in absence of disturbances, given~$\x(0)\in X_N(\lambda)$, the closed-loop system is asymptotically stable.
\end{theorem}

The proof of this Theorem follows from the proof of~\cite[Theorem~1]{manzano2019output}, tailored to the PF problem considered here. First, note that, in absence of uncertainty, the nominal case holds. In other words, the predictions~$\hat \x(j|k)$ match the real system~$\x(k+j),\,\forall j$.

Then, the proof makes use of the following lemmas:

	\begin{lemma}\label{lem:jlessct}
		$\forall \x\in\Omega_\gamma,\, J^*_N(\x)\leq\lambda V_f(\x).$
	\end{lemma}
	
	\begin{proof}
	Summing Eq.~\eqref{eq:deltaVf} over the prediction horizon yields:
	\begin{eqnarray*}
	    &&\sum_{j=0}^{N-1} V_f(\x(j+1|k))-V_f(\x(j|k))\leq \sum_{j=0}^{N-1} -\ell(\x(j|k),\kappa_f(\x(j|k))\\
	    &=& V_f(\x(N|k))-V_f(\x(k))\leq - \sum_{j=0}^{N-1} \ell(\x(j|k),\kappa_f(\x(j|k)).
	\end{eqnarray*}
	Provided that:
	$$J_N^*(\x(k))\leq \sum\limits_{j=0}^{N-1}\ell(\hat \x(j|k),\u(j))+ V_f(\hat\x(N|k))\leq V_f(\x(k)),$$
	\noindent we have that:
	$$ J_N^*(\x(k)) \leq \lambda V_f(\x(k)),$$
	\noindent which proves the lemma.
	\end{proof}
	
	\begin{lemma}
		If~$\x^*(j|k)\notin\Omega_\gamma, \forall j$, then~$J^*_N(\x(k))\geq N\varphi+\lambda\gamma$.
	\end{lemma}
	
	\begin{proof}
	The constant~$\varphi>0$ in Eq.~\eqref{eq_dda} is defined such that:
	$$\forall \x\notin\Omega_\gamma,\,\ell(\x,\u)\geq\varphi.$$
	
	Then, if~$\x^*(j|k)\notin\Omega_\gamma$ for any~$j$,
	$$\sum_{j=0}^{N-1}\ell(\x(j|k),\u)\leq N \varphi,$$
	and~$V_f(\x(N|k))>\gamma$, which leads to the expression in the lemma.
	\end{proof}
	
	\begin{lemma}
		If~$\x^*(N|k)\notin\Omega_\gamma$, then any~$\x^*(j|k)\in\Omega_\gamma,\forall j$.
	\end{lemma}
	
	\begin{proof}
	    Assume that there existed an instant~$j<N$ in which~$\x(j|k)\in\Omega_\gamma$.
	    
	    From Lemma~\ref{lem:jlessct} and optimality we have that~$J_N^*(\x(j|k))\geq\lambda V_f(\x(N|k))$, and provided that~$\x^*(N|k)\notin\Omega_\gamma$, then~$\lambda V_f(\x(N|k))>\lambda\gamma$.
	    Hence~$\lambda V_f(\x(k|k))>\lambda\gamma$. This is a contradiction, which implies that~$\x(j|k)\notin \Omega_\gamma$.
	\end{proof}
	
    In virtue of these lemmas, it can be stated that~$\forall \x\in X_N(\lambda)$, if~$J^*_N(\x(k))\leq N\varphi+\lambda\gamma$, then~$\x(N|k)\in\Omega_\gamma$.
	
	Next, recursive feasibility is proven, defining the future \textit{shifted} sequence:
	\begin{equation}
	    \tilde \u(j|k+1) \equiv \left\lbrace\begin{array}{ll}
	         \u^*(j|k)& j=0,\ldots,N-1  \\
	         \kappa_f(\tilde \x(j|k+1))& j=N, 
	    \end{array}\right.
	\end{equation}
\noindent where~$\tilde \x(j|k+1)$ is the predicted state using~$\u^*(j|k)$ from~$x(k+1)$, for~$j=0,\ldots,N-1$. Note that this sequence is feasible, provided that~$\u^*$ is feasible. The cost obtained using~$\tilde \u(j|k+1)$ from~$x(k+1)$ is denoted~$\tilde J_N(\x(k+1))$. The mismatch between costs is bounded as follows:
\begin{multline}
    \tilde J_N(\x(k+1)) - J_N^*(\x(k)) = -\ell(\x(k),\u^*(k)) \\\nn
    +\Big[ \ell(\tilde \x(N|k+1),\kappa_f(\tilde \x(N|k+1))) + \lambda V_f(\tilde \x(N|k+1)) -\lambda V_f(\x^*(N|k)) \Big],
\end{multline}
where the value among brackets is negative, since~$\x^*(N|k)\in\Omega_\gamma$. By optimality it is known that~$J^*_N(\x(k+1))\leq \tilde J_N(\x(k+1))$ and thus
$$J^*_N(\x(k+1))-J^*_N(\x(k))\leq -\ell(\x(k),\u^*(k)),$$
which implies that the closed-loop system is recursively feasible, i.e., that
	$$ \x(k+1)\in X_N(\lambda).\qed$$
	
\textit{ Stability:} By definition, it is shown that
	\begin{eqnarray}
		J_N^*(\x(k+1))-J_N^*(\x(k))&\leq& -\ell(\x(k),\u(k))\label{eq:l1}\\
		&\leq& -\alpha_x(\|\x\|)-\alpha_u(\|\u-\u_r\|),\nn\\
		J_N^*(\x(k))&\geq& \ell(\x(k),\u(k))\label{eq:l2}\\
		&\geq& \alpha_x(\|\x\|)+\alpha_u(\|\u-\u_r\|),\nn\\
		J_N^*(\x(k))&\leq& \alpha_x(\|\x\|),\label{eq:l3}
	\end{eqnarray}
	\noindent where Eq.~\eqref{eq:l3} holds in virtue of~\cite[Prop.~B.25]{rawlings2017model}, provided that~$X_N(\lambda)$ is closed, and~$J(\cdot)$ is continuous and locally bounded on~$X_N$.

Once nominal stability is proven, the effect of the uncertainty in the disturbance is considered. To do so, input-to-state stability (ISS)~\citep{limon2009input} of the proposed controller is proven, based on the following assumption:

\begin{assumption}\label{ass_v}
	The lateral velocity (sway)~$v$ is upper bounded at all times by some constant~$\overline{v}$:
	\begin{equation}
		|v(k)|\leq \overline{v},\,\forall k.
	\end{equation}
\end{assumption}
Note that we do not require knowledge of the uncertainty bound, only its existence. We find reasonable to assume that the lateral velocity is bounded.

\begin{remark}\label{rem:nu}
	The uncertainty in the model~$f$ is not additive, a priori (cf. Eq.~\eqref{eqPFCErrorDynamics}). However, note that, since~$\x,\u$ in the product with~$v$ are within sine and cosine functions, in virtue of Assumption~\ref{ass_v} we can state that the model mismatch $d(k)$ is:
	\begin{eqnarray}
		d(k)=|f(\x(k),\u(k))-f(\x(k),\u(k),v(k))|\leq 2\overline{v},\nn\\
		\forall \x,\u\in\setU,d\in\mathcal B(\,\overline{v}\,).
	\label{eq_model_mismatch}	
	\end{eqnarray}
\end{remark}

\begin{remark} \label{rem:uc}
	Since~$X_N(\lambda)$ is compact and~$\ell$,~$\kappa_f$, and~$V_f$ are continuous, they are uniformly continuous in~$X_N(\lambda)$.
\end{remark} 

\begin{definition}[ISS stability as defined by~\cite{limon2009input}] A system~$x(k+1)=f(x(k))+d(k)$ is input-to-state stable (ISS) with respect to~$d(k)$ if there exists a~$\mathcal{KL}$-function~$\beta$ and a~$\mathcal K$-function~$\alpha$ such that
\begin{equation}
    \|\x(k)\|\leq \beta(\|\x(0)\|,k)+\sup_{j\in\mathbb I_0^k}\alpha(\|d(j)\|).
\end{equation}

\end{definition}

Hence, the following Theorem states that the closed-loop system is ISS with respect to the model mismatch~$d(k)$.

\begin{theorem}
	If the assumptions of Theorem~\ref{th_nominalstab} and Assumption~\ref{ass_v} hold, the closed-loop system subject to uncertainty~$v(k)$ is ISS with respect to the model mismatch~$d(k)$.
\end{theorem}

The proof of this Theorem holds in virtue of Remarks~\ref{rem:nu} and~\ref{rem:uc}, following from~\cite[Case~C1, Prop.~1]{limon2009input}.


\section{Practical nonlinear model predictive control} \label{secPNMPC}

As stated in Section \ref{sec_intro}, computational cost is in general the main drawback of nonlinear predictive guidance laws, when compared to other LOS-based laws. The NMPC-based law requires to solve the optimization problem stated in Eq.~\eqref{eq_opti} every sampling time. Its complexity is mainly defined by the horizon~$N$ and especially by the nonlinear features of the prediction model indicated in Eq.~\eqref{eq_model} and based on the discretization of Eqs.~\eqref{eqzpunto}~and~\eqref{eqPFCErrorDynamics}.

While short horizons may be suitable for this optimization problem, the nonlinearity of the prediction model, reflected for instance in the trigonometric functions of Eq.~\eqref{eqPFCErrorDynamics}, cannot be avoided. Therefore, the proposed predictive guidance law requires a non-negligible computational capacity, which might not be available in low-cost USV implementations, or might alternatively force the sampling time of the guidance law to be higher than admissible to effectively follow the desired path.

That is the reason why a linearized version of the NMPC-based guidance law presented in Section~\ref{secNMPC} is proposed in this section, inspired in the Practical Nonlinear MPC (PNMPC) technique \citep{plucenio2007practical,Plucenio}. This technique is based on a novel interpretation of the procedure used by traditional linear MPC techniques to compute the predictions, avoiding iterative algorithms, allowing to use the same methods as linear MPC strategies to obtain the control actions, and imposing no constraints on the nonlinear model structure. This technique has been successfully applied to other NMPC-based scheduling and control strategies, obtaining time-efficient implementations of highly nonlinear predictive controllers \citep{bejarano2017optimization, bejarano2019minlp}.

In traditional linear MPC techniques, the vector of predicted outputs $\hat\y$ along horizon $N$ (which might match the vector of system states $\hat\x$, as in this case) can be expressed as a linear function of the vector of future increments on the control inputs $\Delta \u$, where the free response $\y_\mathrm{free}$ and the forced response $\y_\mathrm{forced}$ are explicitly separated, being~$\G$ a constant matrix denominated dynamic matrix of the model, as shown in Eq.~\eqref{eq_DMC_GPC}:
\begin{equation}
	\begin{aligned}
		\hat\y = \y_\mathrm{free} + \y_\mathrm{forced} = \y_\mathrm{free} + \G \cdot \Delta \u. \\
	\end{aligned}
	\label{eq_DMC_GPC}
\end{equation}  
The system model shown in Eq.~\eqref{eq_model} can be also expressed as indicated in Eq.~\eqref{eq_nonlinear}:
\begin{equation}
	\begin{aligned}
		\hat\y =  g(\y_\mathrm{past},\u_\mathrm{past},\Delta \u), \\
	\end{aligned}
	\label{eq_nonlinear}
\end{equation}  
\noindent where the predicted output vector $\hat\y$ turns out to be a given nonlinear function $g(\cdot,\cdot,\cdot)$ of the current and past outputs $\y_\mathrm{past}$, the past control inputs $\u_\mathrm{past}$, and the future increments of the control actions $\Delta \u$. $g$ simply refers to the recursion of the nonlinear function $f$ in Eq.~\eqref{eq_model} along the considered horizon $N$.

If the structure of the linear MPC shown in Eq.~\eqref{eq_DMC_GPC} is intended to be mirrored, the predicted output vector $\hat\y$ can be divided into two parts: the free response $\y_\mathrm{free}$, only due to the current and past outputs $\y_\mathrm{past}$ and the past control inputs $\u_\mathrm{past}$, and the forced response $\y_\mathrm{forced}$, affected by the future increments on the control actions $\Delta \u$. Concerning the free response $\y_\mathrm{free}$, this vector can be easily computed by recursively simulating zero future increments on the control actions to the original nonlinear model expressed in Eq.~\eqref{eq_model} along the horizon $N$, as indicated in Eq.~\eqref{eq_GPNMPC}:
\begin{equation}
	\begin{aligned}
		\y_\mathrm{free} &= g(\y_\mathrm{past},\u_\mathrm{past},\Delta \u=\mathbf 0). \\
	\end{aligned}
	\label{eq_GPNMPC}
\end{equation}  
Regarding the forced response $\y_\mathrm{forced}$, the PNMPC algorithm proposes an approximation based on a first-order linearization of the MacLaurin series, given that it is computed around $\Delta \u = \mathbf 0$, as described in Eq.~\eqref{eq_GPNMPC_2}.
\begin{equation}
	\begin{aligned}
		\y_\mathrm{forced} &\approx \G_\mathrm{PNMPC} \cdot \Delta \u, \\
		\G_\mathrm{PNMPC} &\equiv \dfrac{\partial \hat\y}{\partial \Delta \u} \biggr\rvert_{\Delta \u = \mathbf 0} .
	\end{aligned}
	\label{eq_GPNMPC_2}
\end{equation}  
Therefore, $\G_\mathrm{PNMPC}$ represents the Jacobian matrix, including the gradient of $\hat\y$ with respect to all future increments on the control inputs $\Delta \u$ along the horizon. Eventually, state feedback is applied at every sampling time to avoid offset and close the loop.

Although several numerical algorithms to compute $\G_\mathrm{PNMPC}$ have been presented in the related literature for multiple-input-multiple-output (MIMO) systems \citep{plucenio2007practical,Plucenio,bejarano2017optimization}, in this case the continuous-time equations describing the nonlinear model indicated in Eqs.~\eqref{eqzpunto}~and~\eqref{eqPFCErrorDynamics} can be used to compute analytically the gradients, as shown in Eq.~\eqref{eq_gradient_computation}:
\begin{equation}
	\begin{aligned}
		\frac{\partial \dot{x}_e}{\partial u} \biggr\rvert_0 &= \text{cos}\left(\psi_0 - \phi_{p}(\omega_0)\right),\\
		\frac{\partial \dot{x}_e}{\partial \psi} \biggr\rvert_0 &= - u_0 \, \text{sin}\left(\psi_0 - \phi_{p}(\omega_0)\right) - v_0 \, \text{cos}\left(\psi_0 - \phi_{p}(\omega_0)\right), \\
		\frac{\partial \dot{x}_e}{\partial \utar} \biggr\rvert_0 &= \dot{\phi}_{p_0} \, y_{e_0} - 1 = \dfrac{1}{F(\omega_0)} \dfrac{\partial \phi_p}{\partial \omega}\biggr\rvert_0 \, y_{e_0}  - 1, \\
		\frac{\partial \dot{y}_e}{\partial u} \biggr\rvert_0 &= \text{sin}\left(\psi_0 - \phi_{p}(\omega_0)\right), \\
		\frac{\partial \dot{y}_e}{\partial \psi} \biggr\rvert_0 &= u_0 \, \text{cos}\left(\psi_0 - \phi_{p}(\omega_0)\right) - v_0 \, \text{sin}\left(\psi_0 - \phi_{p}(\omega_0)\right), \\
		\frac{\partial \dot{y}_e}{\partial \utar} \biggr\rvert_0 &=  - \dot{\phi}_{p_0} \, x_{e_0}  = - \dfrac{1}{F(\omega_0)} \dfrac{\partial \phi_p}{\partial \omega}\biggr\rvert_0 \, x_{e_0},  \\
		\frac{\partial \dot{z}}{\partial u} \biggr\rvert_0 &=0, \\
		\frac{\partial \dot{z}}{\partial \psi} \biggr\rvert_0 &= 0, \\
		\frac{\partial \dot{z}}{\partial \utar} \biggr\rvert_0 &= - \dfrac{z(\omega_0)^2}{F(\omega_0)}, \\
	\end{aligned}
	\label{eq_gradient_computation}
\end{equation} 
where the 0 subindex refers to the current instant, applicable to the system states $x_e$, $y_e$, and $z$ (or conversely $\omega$, as expressed in Eq.~\eqref{eq_gradient_computation}), to the manipulated inputs $u$, $\psi$, and $\utar$, and to the measured disturbance $v$.

Then, the Jacobian matrix $\G_\mathrm{PNMPC}$ can be built as shown in Eq.~\eqref{eq_GPNMPC_computation}:
\begin{equation}
	\begin{aligned}
		\G_\mathrm{PNMPC} &= 
		\left[  
		    \begin{matrix}
		        \G_{\mathrm{PNMPC}_{1}} & \mathbf{0} & \mathbf{0} & \cdots & \mathbf{0} \\
		        \G_{\mathrm{PNMPC}_{2}} & \G_{\mathrm{PNMPC}_{1}} & \mathbf{0} & \cdots & \mathbf{0} \\
		        \G_{\mathrm{PNMPC}_{3}} & \G_{\mathrm{PNMPC}_{2}} & \G_{\mathrm{PNMPC}_{1}} & \cdots & \mathbf{0} \\
		        \vdots & \vdots & \vdots & \ddots & \vdots \\
		        \G_{\mathrm{PNMPC}_{N}} & \G_{\mathrm{PNMPC}_{N-1}}& \G_{\mathrm{PNMPC}_{N-2}} & \cdots & \G_{\mathrm{PNMPC}_{1}} \\
		    \end{matrix}
		\right] \in \mathbb{R}^{3N \, \times \, 3N},
	\end{aligned}
	\label{eq_GPNMPC_computation}
\end{equation} 
where $\G_{\mathrm{PNMPC}_{i}} \in \mathbb{R}^{3\,\times\,3}, \; i \in \mathbb{I}_1^N$, can be computed from the gradients indicated in Eq.~\eqref{eq_gradient_computation} as shown in Eq.~\eqref{eq_GPNMPCii_computation}:
\begin{equation}
	\begin{aligned}
		\G_{\mathrm{PNMPC}_{i}} &= i \cdot T_m \cdot
		\left[  
		    \begin{matrix}
		        \frac{\partial \dot{x}_e}{\partial u} \biggr\rvert_0 & \frac{\partial \dot{x}_e}{\partial \psi} \biggr\rvert_0 & \frac{\partial \dot{x}_e}{\partial \utar} \biggr\rvert_0 \\
		        \frac{\partial \dot{y}_e}{\partial u} \biggr\rvert_0 & \frac{\partial \dot{y}_e}{\partial \psi} \biggr\rvert_0 & \frac{\partial \dot{y}_e}{\partial \utar} \biggr\rvert_0 \\
		        \frac{\partial \dot{z}}{\partial u} \biggr\rvert_0 & \frac{\partial \dot{z}}{\partial \psi} \biggr\rvert_0 & \frac{\partial \dot{z}}{\partial \utar} \biggr\rvert_0 \\
		    \end{matrix}
		\right],
	\end{aligned}
	\label{eq_GPNMPCii_computation}
\end{equation} 
being $T_m$ the sampling time. Notice that, since the system to be controlled is nonlinear but time-invariant, the computation of the Jacobian matrix $\G_\mathrm{PNMPC}$ can be simplified by shifting the $\G_{\mathrm{PNMPC}_{i}}$ matrix along the horizon $N$, as shown in Eq.~\eqref{eq_GPNMPC_computation}.

Once the Jacobian matrix $\G_\mathrm{PNMPC}$ is computed at the current instant, together with the free response $\y_\mathrm{free}$, a linear prediction model is available. Therefore, quadratic programming algorithms applied in standard linear MPC formulations can be applied to solve the optimization problem at the current instant, which matches the one detailed in Eq.~\eqref{eq_opti} with the only difference of the linearized prediction model. As a result, the computing time of the optimization is much reduced with respect to the original NMPC-based guidance law described in Section \ref{secNMPC}. At the following sampling time, the linearized model is recomputed again, evaluating the gradients shown in Eq.~\eqref{eq_gradient_computation} at the new instant, and then solving again the optimization problem.


\section{Case study} \label{secSimulation}

In this section, the proposed guidance laws will be applied in simulation to a specific PF problem of the Cybership~II, a~1:70 scale replica of a supply ship for the North Sea, described and identified by~\cite{skjetne2004nonlinear}. The USV dynamics in terms of relative velocities described by~\cite{fossen2011handbook,xia2019improved} have been simulated in order to also consider non-rotational currents as external disturbances. MATLAB\textsuperscript{\textregistered} software has been used to perform all simulations, using a laptop with an Intel\textsuperscript{\textregistered} Core\textsuperscript{\texttrademark} microprocessor i5-7200U CPU @ 2.50 GHz 8 GB RAM.

First, in subsection~\ref{subsec_SGLOS_NMPC}, the proposed NMPC-based guidance law will be compared to other line-of-sight laws existing in the literature, namely the SGLOS law presented by~\cite{wang2019fuzzy}, the ALOS law developed by~\cite{fossen2015line}, and the CLOS law proposed by~\cite{miao2017compound}. Second, once highlighted the advantages of the original NMPC-based guidance over other laws, the reduction of the computational cost given by the PNMPC-based simplified law will be shown in subsection~\ref{subsec_NMPC_PNMPC}, while the performance indices of both strategies will be compared under more realistic conditions.

\subsection{Comparison between the proposed NMPC guidance and other existing laws} \label{subsec_SGLOS_NMPC}

\subsubsection{Description of the case study} \label{subsec_CaseStudy}

In terms of~$\omega$, the desired path in the case study is given by:
\begin{subequations}
	\begin{eqnarray}
		x_p(\omega)&=&1.25\omega+10\sin(2\pi\omega/40)+5,\\
		y_p(\omega)&=&1.75\omega-0.01\omega^2.
	\end{eqnarray}
\end{subequations}
The USV is originally positioned at~$x=\SI{10}{\meter}$,~$y=\SI{10}{\meter}$, while the initial point of the virtual target on the path is assumed to be~\mbox{$\omega=2.5$}. The desired surge velocity $u_{r}$ is \SI{0.15}{\meter\per\second}, while a sampling time of~\SI{1}{\second} is considered for the guidance law.


\subsubsection{Design of the proposed NMPC-based guidance law} \label{subsec_NMPC}

The proposed NMPC-based guidance law is designed as detailed in Section~\ref{secNMPC}, with a quadratic stage cost given by Eq.~\eqref{eq_StageCost}:
\begin{equation} \label{eq_StageCost}
	\ell(\x,\u)=\|\x\|_{\bm{Q}_{\mathrm{NMPC}}}+\|\u-\u_r\|_{\bm{R}_{\mathrm{NMPC}}},
\end{equation}
with diagonal weighting matrices $\bm{Q}_{\mathrm{NMPC}}$ and $\bm{R}_{\mathrm{NMPC}}$ given by:
\begin{subequations}
	\begin{eqnarray}
		\bm{Q}_{\mathrm{NMPC}}&=&\mathrm{diag}(1,1,1\cdot 10^{-5}),\\
		\bm{R}_{\mathrm{NMPC}}&=&\mathrm{diag}(10,1\cdot 10^{-5},1\cdot 10^{-5}).\label{eq_R}
	\end{eqnarray}
\end{subequations}
The control law given by~\cite{wang2019fuzzy} is used as the terminal control law, tuning the parameters such that:
\begin{eqnarray}\label{eq_wang}
	\kappa_f(\x)&=&\begin{bmatrix}
		0.3\sqrt{y_e^2+0.25}\\
		\phi_p(z)-\arctan (y_e/0.5)\\
		0.8 x_e + u\cos(\psi-\phi_p(z))
	\end{bmatrix}.
\end{eqnarray}
The Lyapunov function given by~\cite{wang2019fuzzy} is also used as terminal cost, weighted by the matrix~$\bm{P}_\mathrm{NMPC}$:
\begin{eqnarray}
	V_f(\x)&=&\|\x\|_{\bm{P}_\mathrm{NMPC}}.
\end{eqnarray}

The weighting factor of the terminal cost (recall Eq.~(\ref{eq_opti})) is taken~$\lambda=1.1$, and the weight $\bm{P}_\mathrm{NMPC}$ is computed following standard procedures of MPC (see~\cite{rawlings2017model}), using a linearized state-space model around the equilibrium point given by~\mbox{$\x=(0,0,1 \cdot 10^{-2})$}~and~\mbox{$\u=(0.1 \cdot u_{r},\phi_p, 0.1 \cdot u_{r})$}, applied to the desired path, previously defined.

The constraints on the inputs are given by Eq.~\eqref{eq:constraints}, with parameters~$\epsilon$=\SI{0.01}{\meter\per\second}, $\bar u$=\SI{0.225}{\meter\per\second}, $\bar{u}_\mathrm{tar}$=\SI{0.75}{\meter\per\second}, $\overline{\delta u}$=\SI{0.05}{\meter\per\second}, and $\overline{\delta \psi}=\pi/4 \; \text{rad}$. The reference vector for the control actions is given by~$\u_r=(u_{r},0,u_{r})$, although in practice, the weights chosen in~$\bm{R}_{\mathrm{NMPC}}$ (cf. Eq.~\eqref{eq_R}) imply a reference only on the surge. The prediction and control horizon $N$ is set to 3 samples and the interior-point algorithm is used as the numerical optimization procedure to solve the nonlinear problem.


\subsubsection{Results} \label{subsec_SGLOS_NMPC_Transient}

\begin{figure}[t]
	\begin{center}
		\includegraphics[width=\columnwidth]{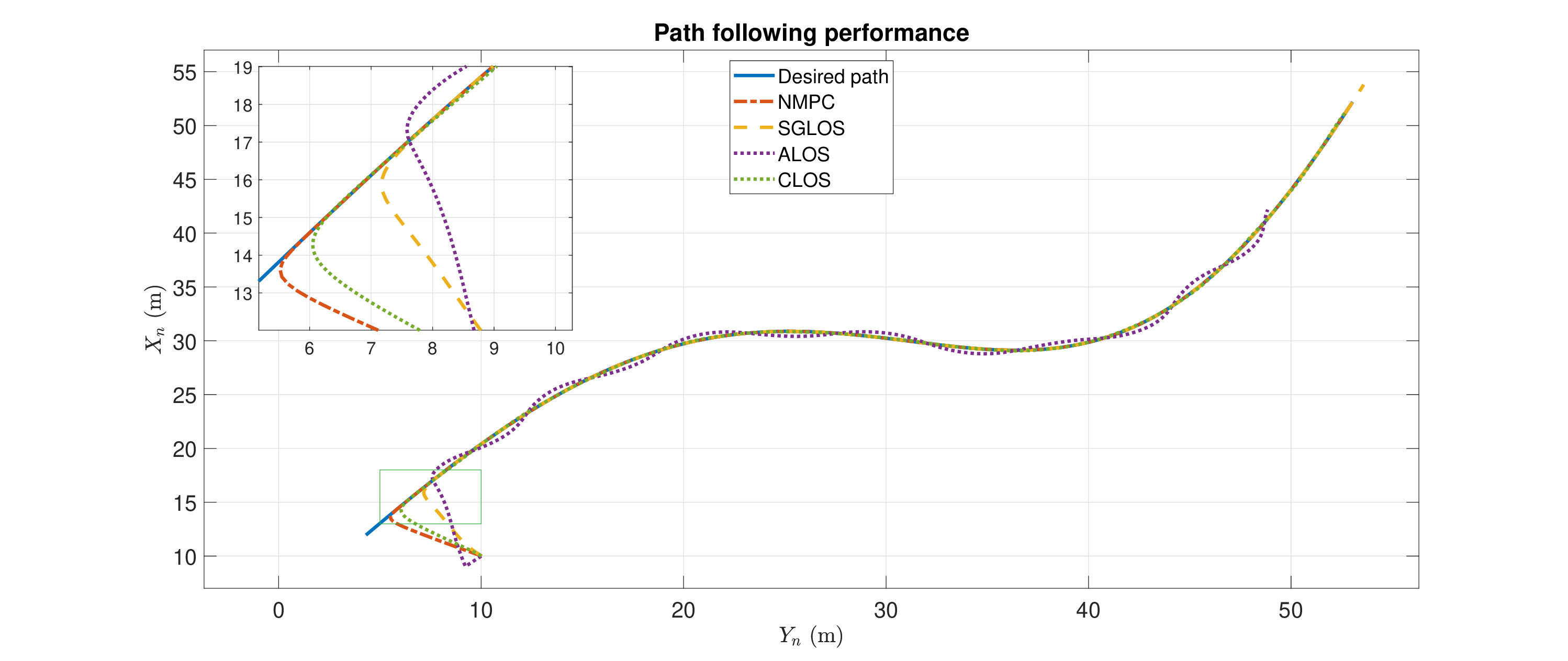} 
		\caption{Desired path and comparison between performances of the NMPC with respect to existing LOS laws}
		\label{fig_tray_Wang_Transient}
	\end{center}
\end{figure}

A closed-loop simulation with the proposed NMPC-based guidance law is shown in Fig.~\ref{fig_tray_Wang_Transient}, where the sway disturbance is a sinusoidal signal with amplitude \SI{0.15}{\meter\per\second} and period \SI{60}{\second}. The desired path is represented in blue and the performance of the proposed law is the orange dashed-dotted line.

This performance is compared to other three laws existing in the literature. In these comparisons, the control actions are forced to meet the bounds detailed in subsection~\ref{subsec_NMPC}. Such guidance laws have been tuned to achieve fast convergence to the desired path while ensuring a smooth behaviour. The SGLOS law proposed by~\cite{wang2019fuzzy} is shown with the dashed yellow line. Its parameters are $\Delta$ = \SI{0.5}{\meter}, $k_1$ = \SI{0.3}{\per\second}, and $k_2$ = \SI{0.8}{\per\second}. The ALOS law by~\cite{fossen2015line} is represented by the dotted purple line, with parameters~$\Delta$ = \SI{0.5}{\meter} and~$\gamma$ = 0.003. Finally, the CLOS by~\cite{miao2017compound} is shown in green dotted line, with parameters~$\Delta$ = \SI{0.5}{\meter} and~$k_1$ = \SI{0.8}{\per\second}. All these parameters are named according to the original notation by their authors,~\cite{wang2019fuzzy},~\cite{fossen2015line}, and~\cite{miao2017compound}, respectively.

It can be observed that the NMPC-based law approaches the path in a more direct way, incurring into smaller PF errors. Once the path is reached, all the controllers exhibit a similar behaviour, except for the ALOS.

Note that a more conservative tuning of the existing LOS laws, trying not to violate the input constraints, could have been considered, at the expense of reaching the neighbourhood of the origin more slowly. Moreover, it is important to remark that tuning the existing LOS laws is not a trivial problem. Furthermore, unlike the NMPC-based guidance law, it is not possible to theoretically impose any constraint on the existing LOS laws, neither on the control actions nor on their increments.

\begin{figure}[h]
	\begin{center}
		\includegraphics[width=\columnwidth]{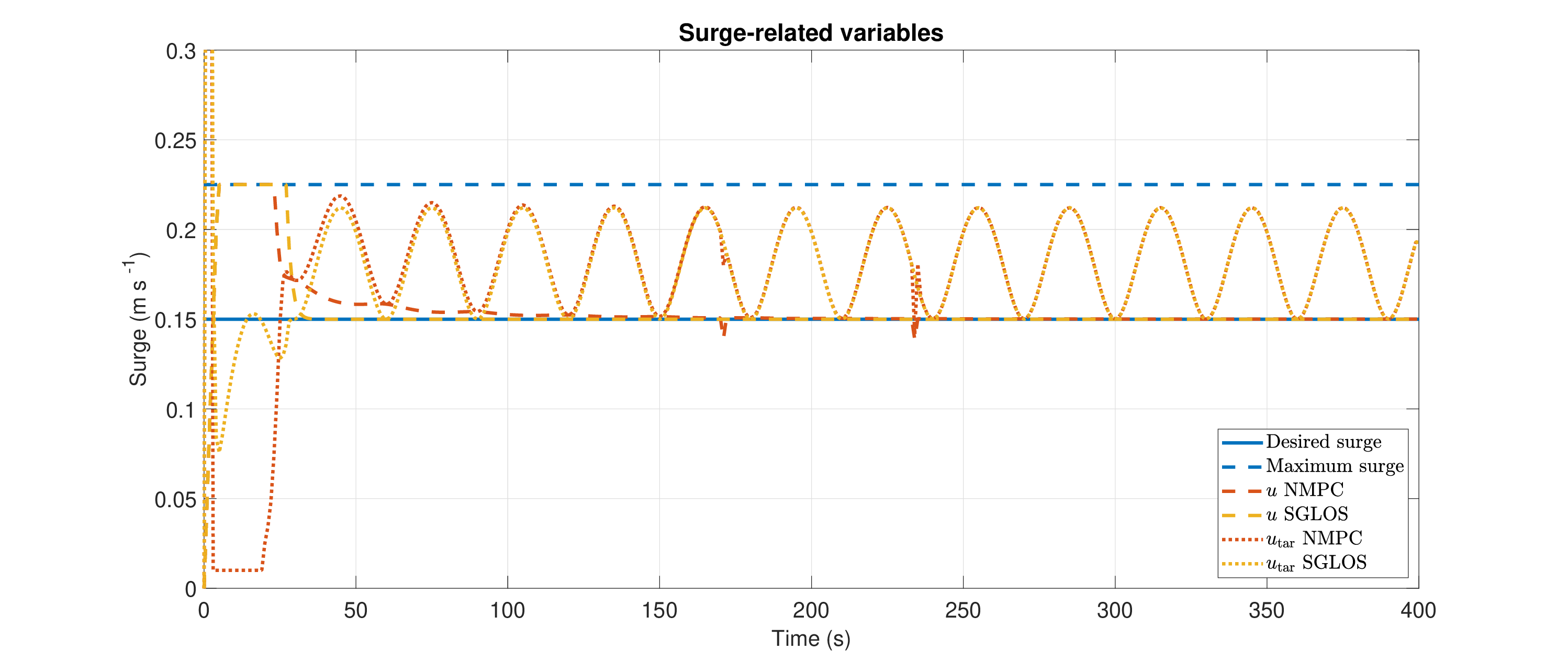}
		\caption{Surge-related variables of both the SGLOS and the NMPC-based guidance laws} 
		\label{fig_surge_Wang_Transient}
	\end{center}
\end{figure}

\begin{figure}[!t]
	\begin{center}
		\includegraphics[width=.8\columnwidth]{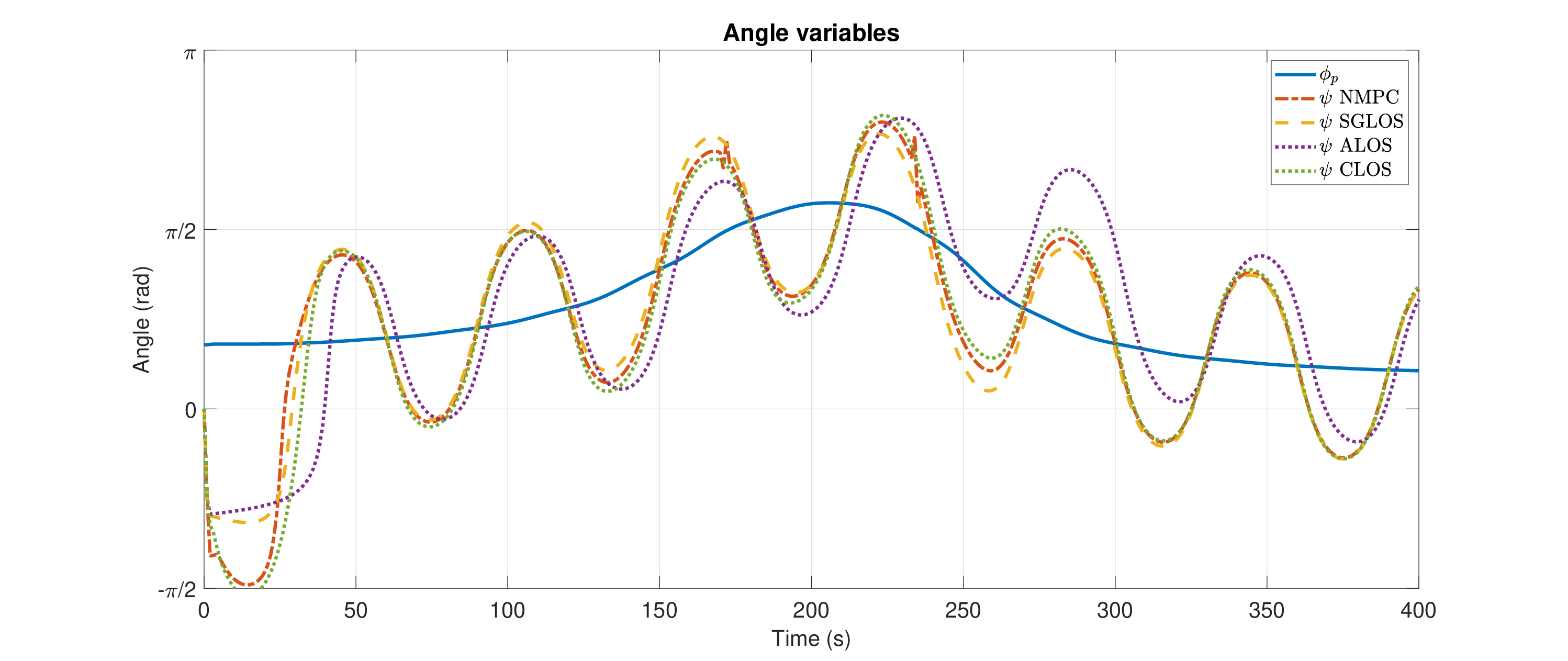} 
		\caption{Angle variables of existing LOS and the NMPC-based guidance laws}
		\label{fig_heading_Wang_Transient}
	\end{center}
\end{figure}

The control actions of the closed-loop system are shown in Fig.~\ref{fig_surge_Wang_Transient}~and~\ref{fig_heading_Wang_Transient}. The surge (dashed orange line) and the virtual target velocity (dotted orange line) for the proposed guidance law are shown in Fig.~\ref{fig_surge_Wang_Transient}, together with the desired value (solid blue line), the maximum allowed surge (dashed blue line), and finally the surge (dashed yellow line) and the virtual target velocity (dotted yellow line) for the SGLOS law. The surge in the ALOS and CLOS strategies is not controllable, as they assume a steady forward speed of the vessel. It can be noted that the constraints on the maximum surge and the maximum allowed variations of the surge along time are met, thanks to the constraint awareness of the NMPC framework, while the SGLOS control actions have to be externally bounded.

In Fig.~\ref{fig_heading_Wang_Transient}, the heading angle is shown, together with the path angle~$\phi_p$ (blue line). Furthermore, the heading angles corresponding to the other LOS laws are represented. Notice that, for all the controllers, the virtual target velocity (when used) and heading angle present an oscillating behaviour, due to the imposed oscillating profile of the disturbance~$v$, which is rejected by such control actions. 

\begin{figure}[!t]
	\begin{center}
		\includegraphics[width=.8\columnwidth]{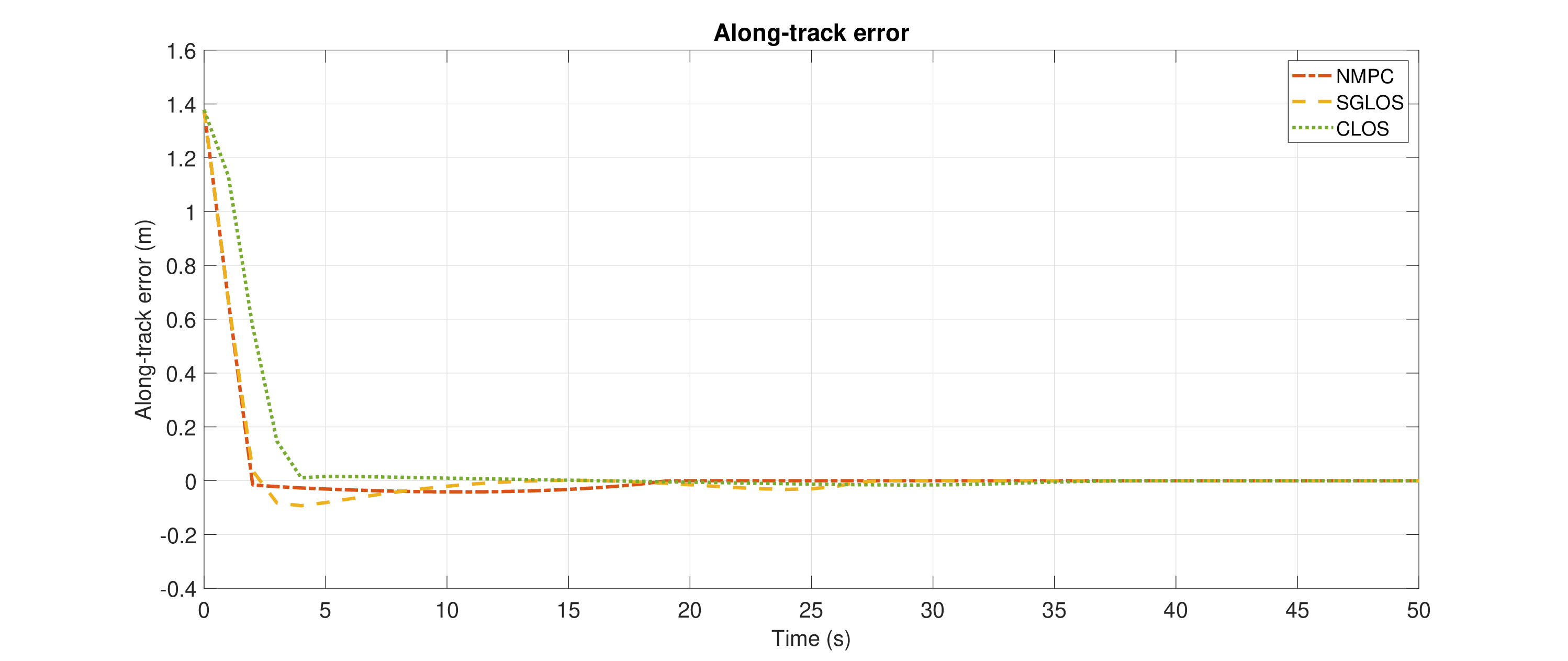} 
		\caption{Comparison of the along-track error for the SGLOS, CLOS, and the NMPC-based guidance laws}
		\label{fig_xe_Wang_Transient}
	\end{center}
\end{figure}

\clearpage

\begin{figure}[!t]
	\begin{center}
		\includegraphics[width=.8\columnwidth]{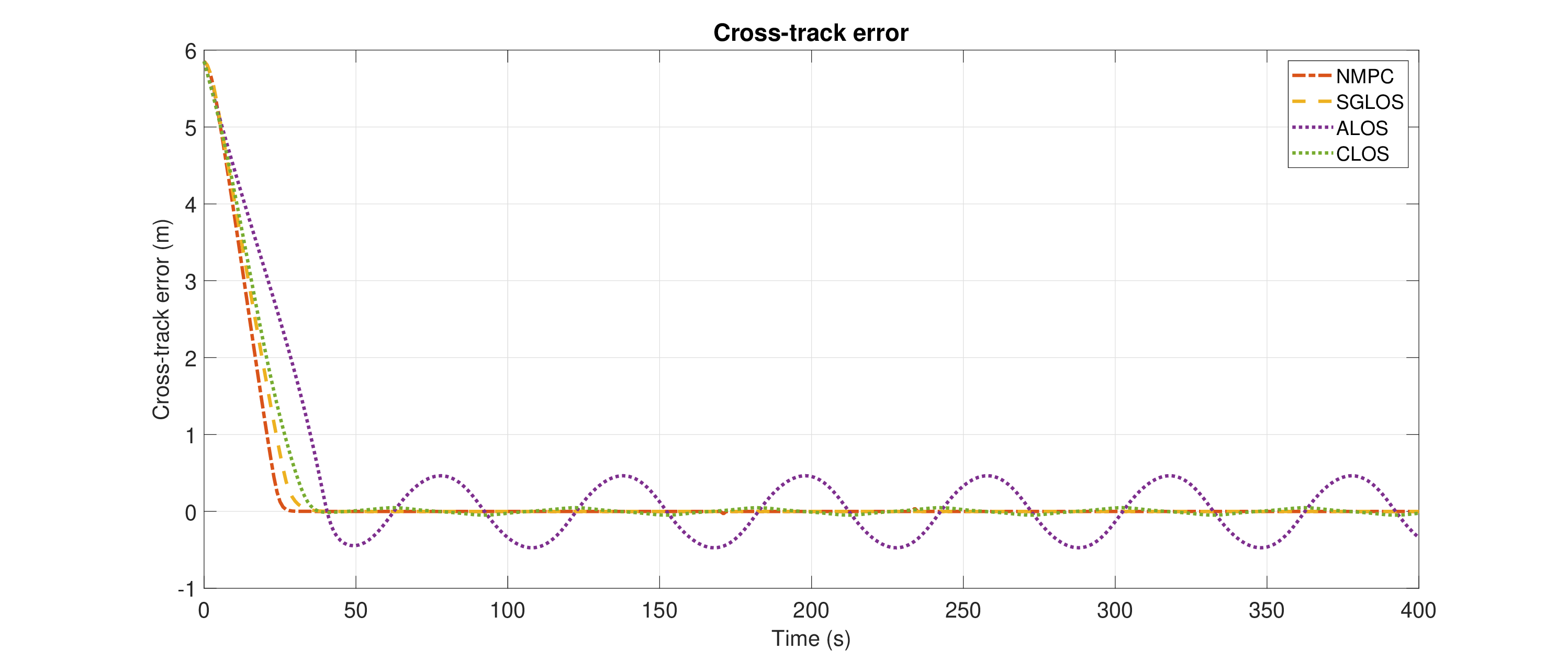} 
		\caption{Comparison of the cross-track error for the LOS and the NMPC-based guidance laws}
		\label{fig_ye_Wang_Transient}
	\end{center}
\end{figure} 

The along- and cross-track errors are represented in Fig.~\ref{fig_xe_Wang_Transient}~and~\ref{fig_ye_Wang_Transient}, respectively. These figures compare the error obtained with the proposed NMPC-based law (dashed-dotted orange line) and the existing LOS laws with forced input constraints. In case of the along-track error, which is a virtual signal, the ALOS technique does not consider a virtual target point, and hence~$x_e$ does not exist. The NMPC drives such error to zero faster than the CLOS and the SGLOS, which oscillate around the origin and stabilize slower, as shown in Fig.~\ref{fig_xe_Wang_Transient}. The NMPC law is also faster when driving the cross-track error to zero, compared to the mentioned existing LOS laws, as shown in Fig.~\ref{fig_ye_Wang_Transient}.

Given that in the simulations presented in this subsection the sway~$v$ is bounded and its upper bound is~$\overline{v}$ = \SI{0.15}{\meter\per\second}, please notice that the input-to-state stability of the proposed NMPC controller is theoretically ensured with respect to the model mismatch~$d$, which turns out to be bounded by~$2\overline{v}$, according to Eq.~\eqref{eq_model_mismatch}. 


\subsection{Comparison between the NMPC-based guidance law and the PNMPC-based simplified law} \label{subsec_NMPC_PNMPC}

Although an ideally perfect low-level control is usually assumed in the PF literature (i.e., ideal tracking of the set points computed by the guidance law, given the cascade control structure), several uncertainties and considerations will be introduced in this subsection to simulate more realistic conditions under which the whole control strategy is likely to operate in real situations. Namely, uncertainties caused by different sampling times between the high and low-level controllers, described in subsection~\ref{subsec_dynamics}, low-level closed-loop dynamics, detailed in subsection~\ref{LowLevelDynamics}, and frequency-varying disturbances, indicated in subsection~\ref{FrequencyVaryingDisturbances}, are simultaneously considered. The case study and the parameters of the NMPC-based guidance law are the same as already detailed in subsection~\ref{subsec_SGLOS_NMPC}. The simulation results of both the original NMPC-based law and the simplified PNMPC-based one are compared in subsection~\ref{subsec_Results}.

\subsubsection{Separate dynamics} \label{subsec_dynamics}

Considering that the proposed NMPC-based law uses a receding horizon technique (that is, an optimization problem must be solved every time step), the choice of the sampling time is not trivial. On the one hand, if a too small sampling time is chosen, the time that the controller takes to obtain the solution of the optimization problem may be longer than the sampling time itself. On the other hand, if the sampling time is very large, the frequency with which the controller changes the references of the surge and heading will be very low, so optimality could be affected. Therefore, selecting the sampling time is always a trade-off between optimality and feasibility. In this case study, a sampling time of~\SI{1}{\second} has been used for the guidance law, considering that the computing power of the on-board computer in charge of implementing the motion control could be limited in the case of low-cost USVs. 

The sampling time of the low-level control usually has more severe time constraints as that of the guidance law, to ensure the suitable tracking of the set points provided by the latter. Thus, assuming a cascade control structure, the sampling time of the low-level controller is reduced one order of magnitude. In this case study, the model of the low-level control block and the system described in Eq.~\eqref{eq_model} are integrated considering an integration step equal to the sampling time of the low-level controller, to simulate the difference between those sampling times. Specifically, the value used for this integration step is~\SI{100}{\milli\second}.

Notice that considering different sampling times introduces errors in the high-level controller predictions, because such predictions are computed using the model with an integration step of~\SI{1}{\second} and with the value of the disturbances sampled at that time. However, the simulation is performed with an integration step of~\SI{100}{\milli\second} with the disturbances sampled at that shorter time. This means that, from the high-level controller point of view, the disturbance signal is constant during all its higher sampling time, but in the simulation, the disturbance varies within that interval. The uncertainty due to the difference between high- and low-level sampling times is reflected in the prediction errors.

\subsubsection{Low-level closed-loop dynamics} \label{LowLevelDynamics}

\begin{figure}[t]
\centering
\begin{minipage}{.5\textwidth}
  \centering
  \includegraphics[width=\linewidth]{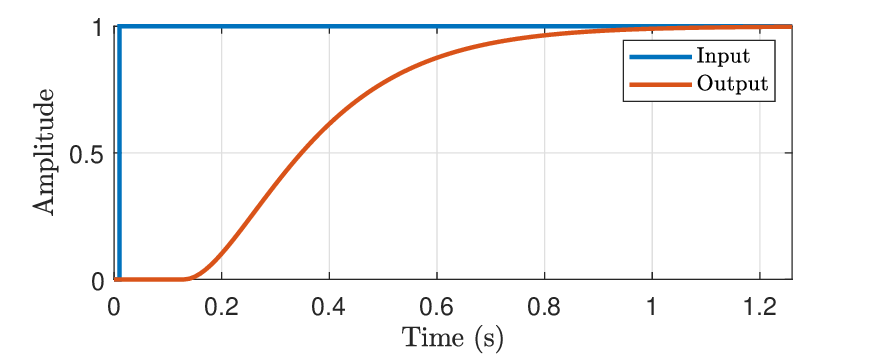}
  \captionof{figure}{Low-level closed-loop dynamics}
  \label{fig_filter}
\end{minipage}%
\begin{minipage}{.5\textwidth}
  \centering
  \includegraphics[width=.9\linewidth]{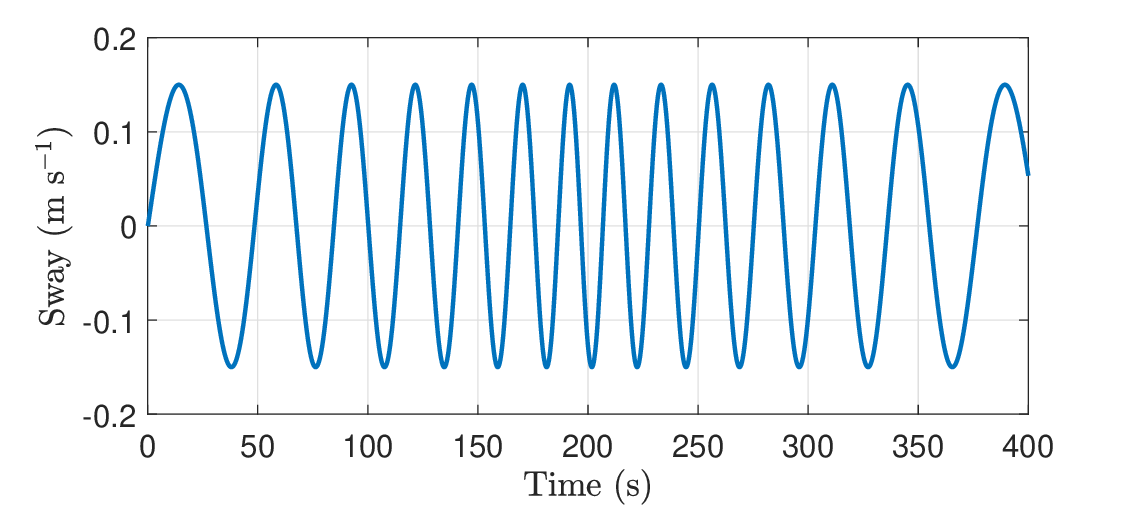}
  \captionof{figure}{Frequency-varying profile for the sway}
  \label{fig_sway}
\end{minipage}
\end{figure}

From the guidance law point of view, the low-level controller is usually assumed to perfectly track the references. In addition to the uncertainty described in subsection~\ref{subsec_dynamics}, the uncertainty generated by the dynamics of the actuators and the low-level controller are intended to be considered. In practice, the actuator capabilities must be taken into account, since actual motors and rudders do not provide instantaneous response, causing the surge and heading angle to take a while to track the set points provided by the guidance law. In this work, the modelling of these dynamics is performed by filtering the reference signals (surge and heading angle) provided by the guidance law. The selected filter is a second-order transfer function with real double poles at $s=-7.6923$ and with an output delay of~\SI{0.13}{\second}. Fig.~\ref{fig_filter} shows the filtered output for a unit step input.

Notice that, when filtering the surge and heading references, the set points provided to the low-level controller are different to the expected set points of the high-level controller, causing discrepancies between the predicted and the simulated behaviour. This is another source of uncertainty, which together with the previous one, intends to emulate a more realistic behaviour.

\subsubsection{Frequency-varying disturbance} \label{FrequencyVaryingDisturbances}

As stated in Section~\ref{secModellingProblemFormulation}, the disturbances affecting the system can be due to different external sources, such as wind, waves, and currents. As explained in Section~\ref{secNMPC}, the effect of all these disturbance sources is gathered in the non-measurable sway velocity $v$. 

In order to study different unfavorable situations for the proposed controller, a frequency-varying profile has been introduced as for the disturbance~$v$. Fig.~\ref{fig_sway} shows the proposed profile for the sway along time. It is composed of two mirrored chirp signals, with amplitude~\SI{0.15}{\meter\per\second} (matching the value of the desired surge $u_{r}$), varying from~1/60 Hz to~1/30 Hz at~\SI{200}{\second}, and vice versa.

\subsubsection{Results} \label{subsec_Results}

\begin{figure}[!ht]
	\begin{center}
		\includegraphics[width=\columnwidth]{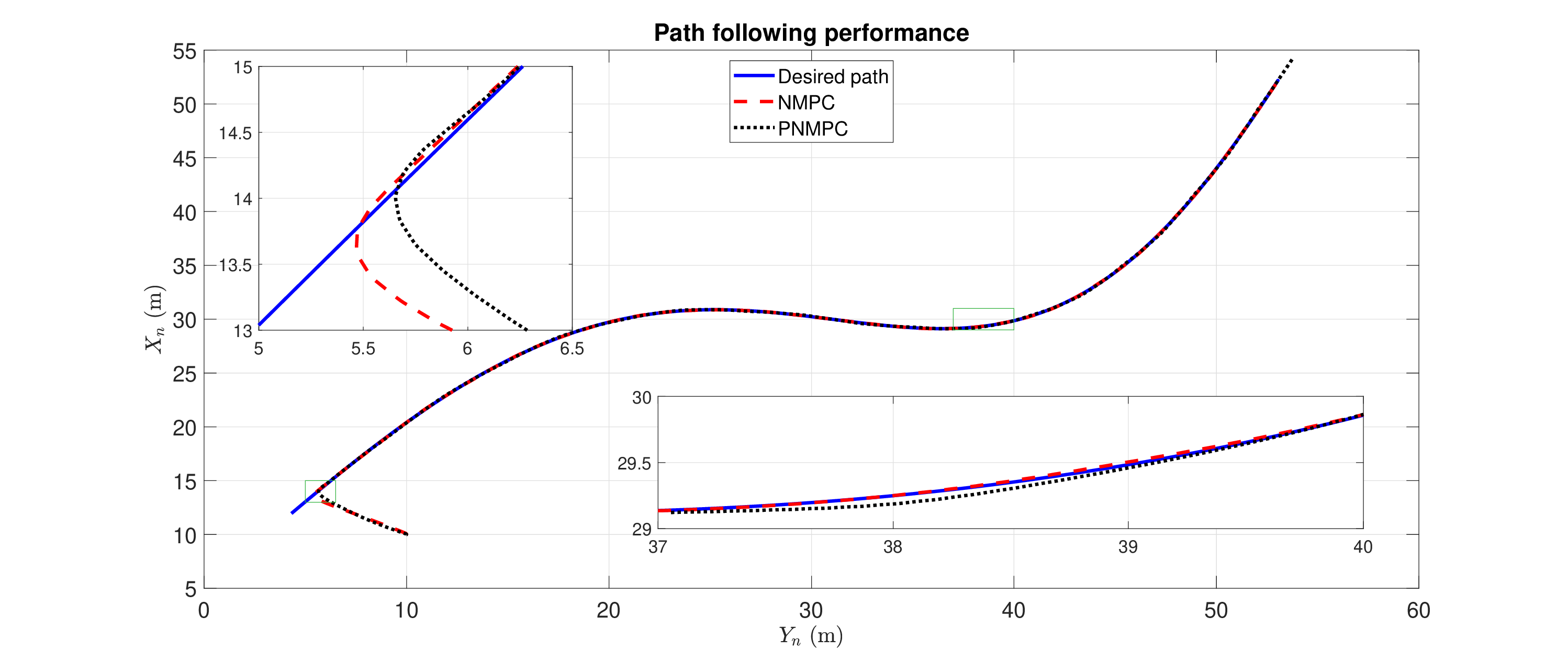}  
		\caption{Path following performances of the NMPC- and PNMPC-based guidance laws} 
		\label{fig_tray_PNMPC}
	\end{center}
\end{figure}

A closed-loop simulation with both the original NMPC-based guidance law and the simplified PNMPC-based law is shown in Fig.~\ref{fig_tray_PNMPC}, where the desired path is represented in blue and the performance of the proposed predictive laws are the dashed red and the dotted black lines respectively. 

In the upper left corner, Fig.~\ref{fig_tray_PNMPC} also shows in detail how the proposed controllers approach the path. Furthermore, the effect of disturbances can be appreciated in the lower right corner of the figure. Notice that both predictive strategies have a similar path following performance, although the complex NMPC-based law approaches the path faster than the simplified PNMPC-based law.

\begin{figure}[!ht]
	\begin{center}
		\includegraphics[width=\columnwidth]{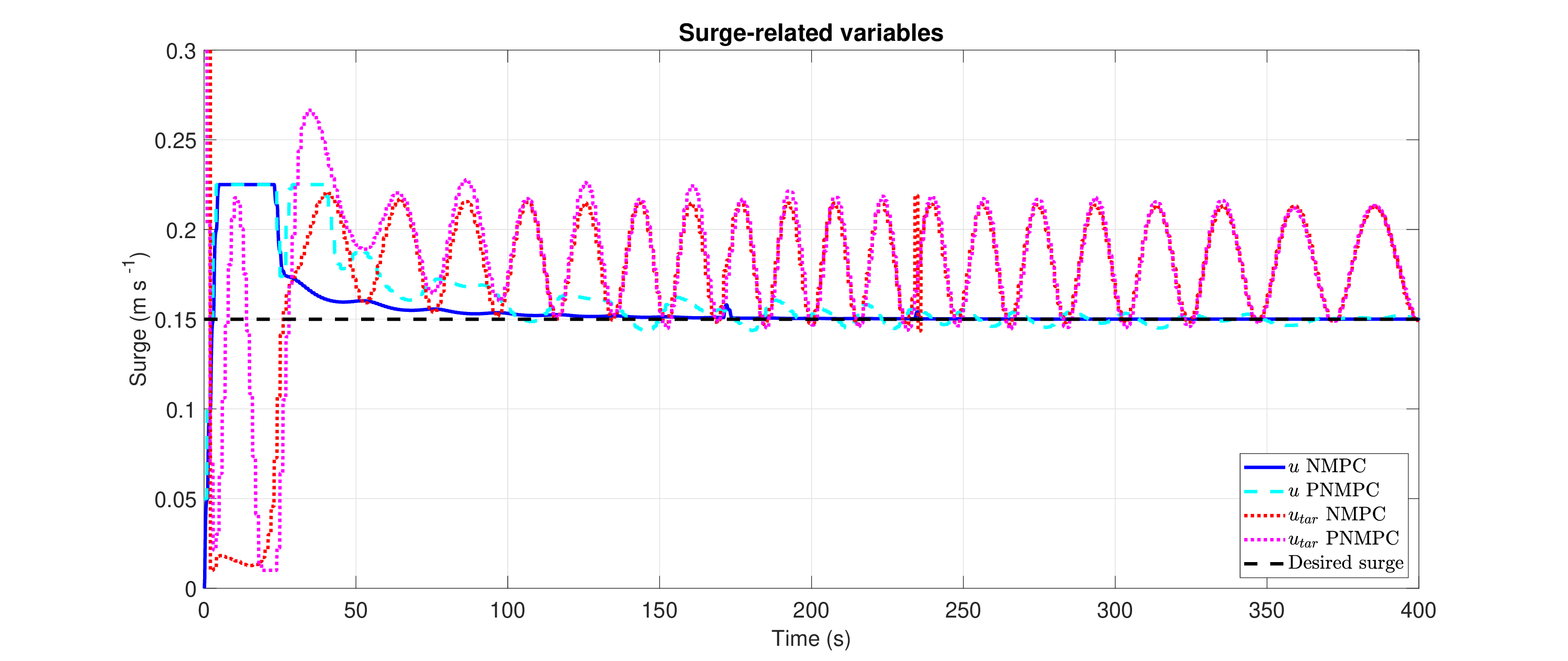} 
		\caption{Surge-related variables of the NMPC- and PNMPC-based guidance laws}
		\label{fig_surge_PNMPC}
	\end{center}
\end{figure}

\begin{figure}[!ht]
	\begin{center}
		\includegraphics[width=\columnwidth]{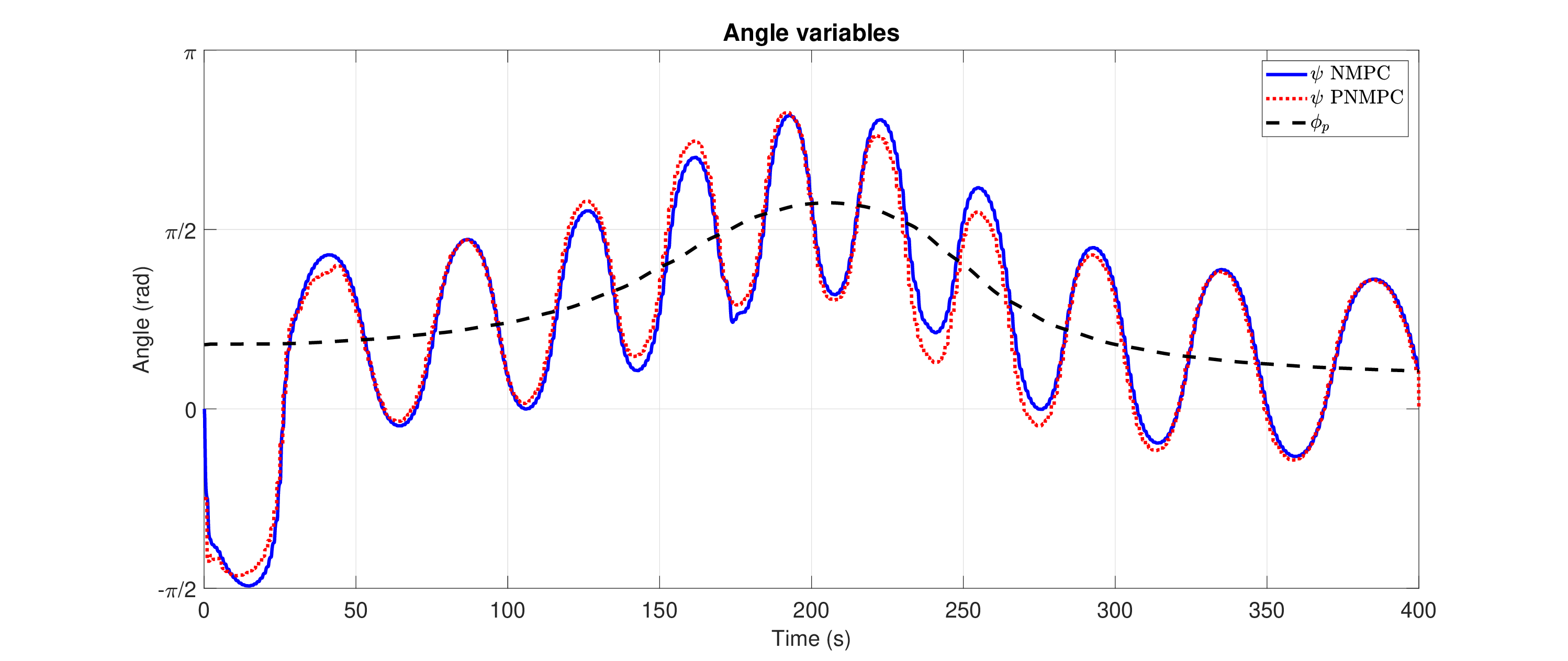}
		\caption{Angle variables of the NMPC- and PNMPC-based guidance laws}
		\label{fig_heading_PNMPC}
	\end{center}
\end{figure}

The set points of the surge and virtual target velocity for both strategies are shown in Fig.~\ref{fig_surge_PNMPC}, as well as the desired surge, while Fig.~\ref{fig_heading_PNMPC} represents the set points of the heading angle, together with the path angle $\phi_p$. 

Notice how the constraints on the maximum values and maximum allowed variations of the surge and heading along time are met by both predictive strategies, thanks to the constraint awareness of the predictive framework. Notice also that both the virtual target velocity and the heading angle present an oscillating behaviour, whose frequency varies due to the sway. Since the surge must be kept as close as possible to the desired value, once that the USV is close to the desired path, the controller uses the remaining control actions ($\utar$ and $\psi$) to reject the disturbance caused by the sway $v$. Notice that, when the disturbance has the highest frequency (around~\SI{200}{\second}), the deviations of the surge with respect to the desired value are greater, since the constraints imposed on the increment of the control actions make insufficient the feasible oscillating response to reject completely the disturbance. 

\begin{figure}[!ht]
	\begin{center}
		\includegraphics[width=\columnwidth]{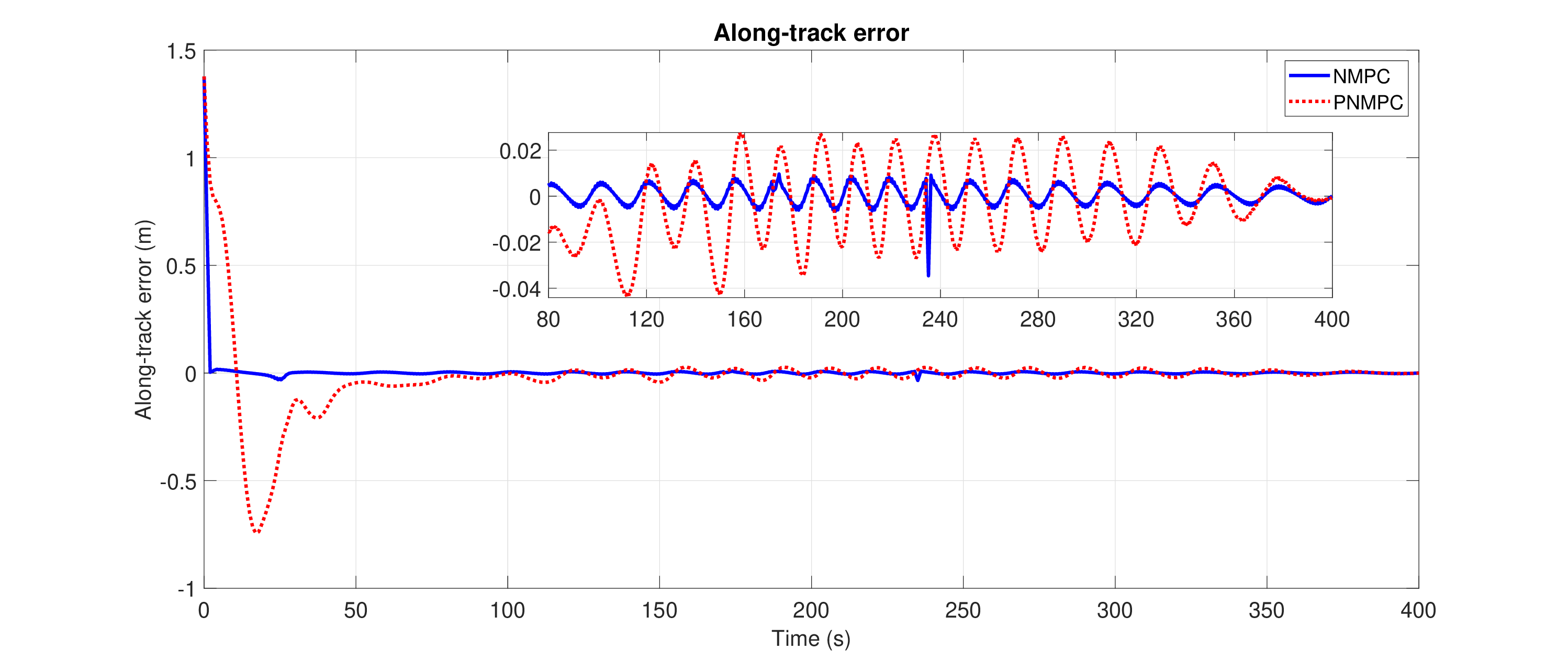} 
		\caption{Comparison of the along-track error for the NMPC- and PNMPC-based guidance laws}
		\label{fig_xe_PNMPC}
	\end{center}
\end{figure}

\begin{figure}[!ht]
	\begin{center}
		\includegraphics[width=\columnwidth]{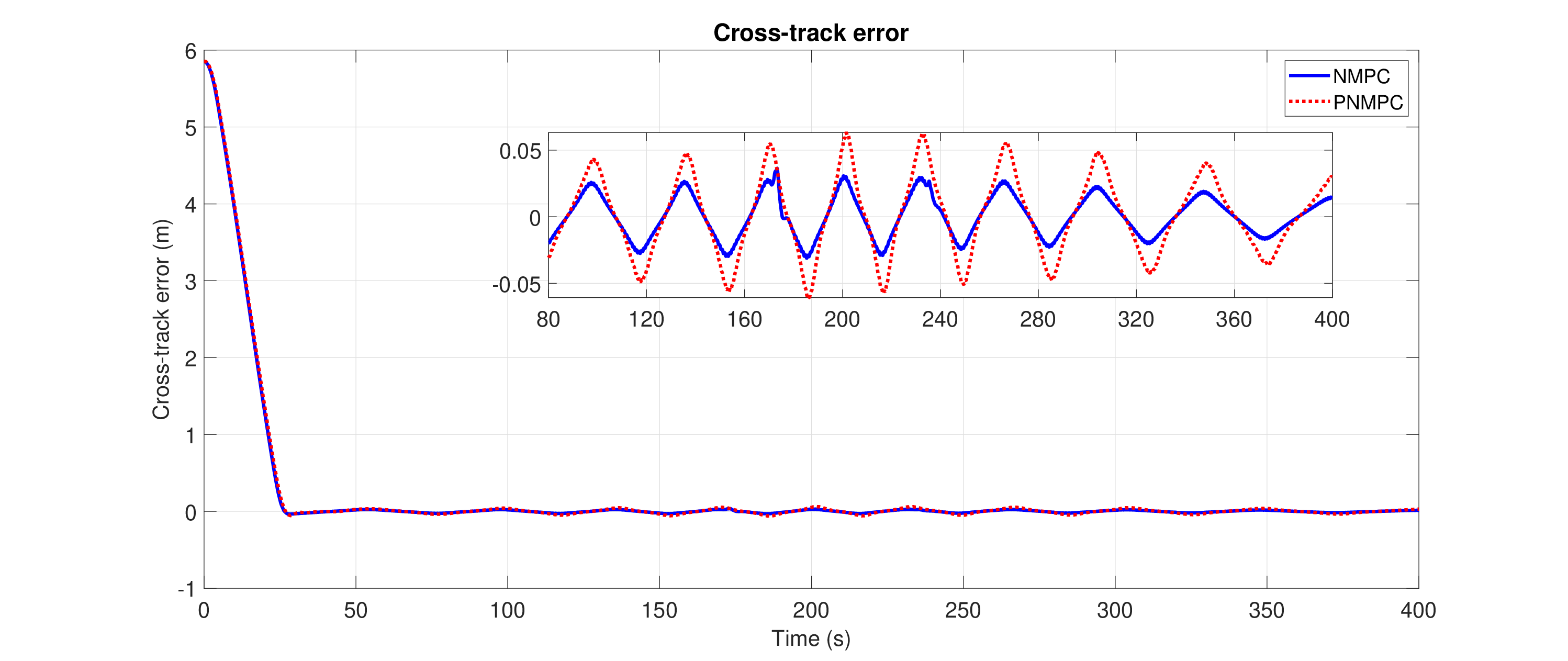} 
		\caption{Comparison of the cross-track error for the NMPC- and PNMPC-based guidance laws}
		\label{fig_ye_PNMPC}
	\end{center}
\end{figure}

The PF errors are shown in Fig.~\ref{fig_xe_PNMPC}~and~\ref{fig_ye_PNMPC}, respectively, with detailed portions around the origin in both figures. Both strategies present again similar performance, although the NMPC-based law incurs in lower PF errors than the PNMPC-based law. 

\begin{figure}[!t]
	\begin{center}
		\includegraphics[width=0.6\columnwidth]{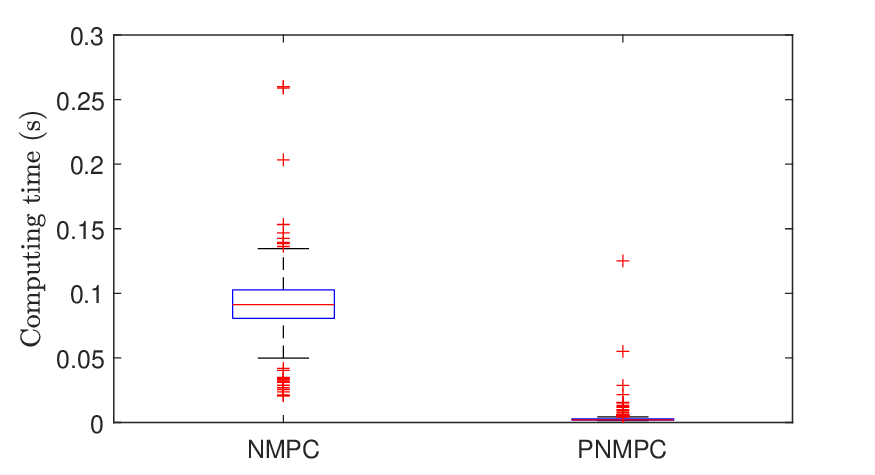}
		\caption{Boxplot of the computing times of the NMPC- and PNMPC-based guidance laws}
		\label{fig_time}
	\end{center}
\end{figure}

Finally, the computing times of both strategies are compared in Fig.~\ref{fig_time}. As described in subsection~\ref{subsec_dynamics}, in this case study the sampling time of the high-level controller is set to~\SI{1}{\second}. Thus, this sampling time is a computational constraint to solve the optimization problem presented in Eq.~\eqref{eq_opti}, for both predictive strategies. Fig.~\ref{fig_time} compares the computing times for the 400 iterations of the optimization problem.

It can be observed that, despite all the iterations being solved satisfying the sampling time constraint in both strategies, the computing times of the PNMPC-based controller are much lower than those of the original NMPC-based law, achieving an average reduction to the 3.87\% of the NMPC computing time. Therefore, the simplified PNMPC-based guidance law is shown to reduce the computational cost of the original NMPC-based guidance law while providing a similar performance in terms of path following, which makes it more suitable to be implemented on on-board computers in the case of low-cost USV implementations. 


\section{Conclusions and future work} \label{secConclusions}

This work has proposed a nonlinear model predictive control-based guidance law for path following problems applied to unmanned surface vehicles. In addition to enable the application of predictive strategies to the low-level controller, through the computation of future references for the surge and heading angle, this guidance law overcomes some drawbacks of other line-of-sight-based guidance laws, which cannot consider constraints or use the a priori knowledge of the desired path and surge. The stability and robustness of the proposed predictive guidance law have been theoretically shown, whereas indications about parameter tuning of the predictive strategy have been also given. 

However, while non-predictive guidance laws are usually effective enough for PF problems and very fast to implement, the proposed strategy involves solving a highly nonlinear optimization problem at every sampling time, thus requiring a non-negligible computational capacity. A practical predictive strategy has been applied to generate a simplified version of the proposed law, by linearizing the model along the trajectory. This allows to apply quadratic programming algorithms to solve the optimization problem, giving rise to a faster implementation.

The effectiveness and superiority of the proposed guidance law over various recently proposed non-predictive strategies has been illustrated in simulation, while the linearized version of the proposed predictive law has shown to provide an average reduction to the 3.87\% of the computing time of the NMPC optimization problem.

As future works, the extension of the proposed strategy to the output constrained case for obstacle avoidance is intended, as well as integrating the guidance law with a predictive low-level controller. In addition, a deeper analysis on the robustness performance of the proposed predictive controller under a variety of dynamical uncertainty sources, such as those described in subsection~\ref{subsec_NMPC_PNMPC}, will be addressed. Moreover, the proposed algorithms are intended to be applied to actual USVs that are currently being built at the laboratory of Universidad Loyola Andaluc{\'i}a\footnote{\url{https://eventos.uloyola.es/59990/section/28306/laboratorios-de-optimizacion-y-control-de} \url{- sistemas -distribuidos.html}}.


\section*{Acknowledgements} \label{secAcknowledgments}

The authors would like to acknowledge Spanish AECID (YPACARAI project, reference 2018/ACDE/000773), Junta de Andaluc{\'\i}a (project reference PY18-RE-0009) and the Agencia Estatal de Investigación (AEI)-Spain under Grant PID2019-106212RB-C41/AEI/10.13039/501100011033, for partially funding this work.


\bibliographystyle{cas-model2-names}

\bibliography{bibliography}             

\newpage

\bio{GBP_photo}
Guillermo Bejarano completed his Ph.D. in Automation, Electronics, and Telecommunications at the University of Seville, Spain. After finishing his Ph.D., he moved to the Engineering School at the Universidad Loyola Andaluc\'ia, Seville, Spain. He is currently a Researcher at the Universidad Loyola Andaluc\'ia, Spain, with the Department of Engineering. His main research interests include smart agriculture and motion control of autonomous marine vehicles.
\endbio
\vspace*{0.8cm}

\bio{pepe}
J.M. Manzano received his MSc degree in Industrial Engineering and his Ph.D. in Automation Engineering from the University of Seville in 2016 and 2020, respectively. He carried out his research in the Department of Systems and Automation of the University of Seville, and he was a visiting researcher at the University of Oxford in 2019. He currently holds a Lecturer position at the Engineering Department of the Universidad Loyola Andaluc\'ia, Spain. His research interests merge nonlinear model predictive control and data-based learning algorithms. He focuses on the study of stability and robustness properties of MPCs whose models are inferred from observed data, using machine learning techniques.
\endbio

\bio{JR_loyola}
Jos\'e Ram\'on Salvador received his M.Eng. in telecommunication engineering from the University of Granada, Spain, and completed his Ph.D. in Automation, Electronics, and Telecommunications at the University of Seville in 2019. He has been a visiting researcher at the Politecnico di Milano, Italy. Currently, he is an Assistant Professor with the Department of Engineering at the Universidad Loyola Andaluc\'ia, Spain. His main research interests include model predictive control and data-based predictive control.
\endbio

\bio{dani}
Daniel Limon received the M.Eng. and Ph.D. degrees in electrical engineering from the University of Seville, Spain, in 1996 and 2002, respectively.
From 1999 to 2007, he was an Assistant Professor with the Departamento de Ingenier\'ia de Sistemas y Autom\'atica, University of Seville, from 2007 to 2017, Associate Professor and since 2017, a Full Professor in the same Department. He has been a visiting researcher at the University of Cambridge and the Mitsubishi Electric Research Labs in 2016 and 2018, respectively. Dr. Limon has been a Keynote Speaker at the International Workshop on Assessment and Future Directions of Nonlinear Model Predictive Control in 2008 and Semiplenary Lecturer at the IFAC Conference on Nonlinear Model Predictive Control in 2012. He has been the Chair of the fifth IFAC Conference on Nonlinear Model Predictive Control (2015). His current research interests include model predictive control, stability and robustness analysis, tracking control, and data-based control.
\endbio

\end{document}